# Machine learning to explore high-entropy alloys with desired enthalpy for room-temperature hydrogen storage: Prediction of density functional theory and experimental data

Shivam Dangwal[1,2], Yuji Ikeda[3], Blazej Grabowski[3] and Kaveh Edalati[1,2,*]

[1] WPI, International Institute for Carbon-Neutral Energy Research (WPI-I2CNER), Kyushu University, Fukuoka, Japan
[2] Department of Automotive Science, Graduate School of Integrated Frontier Sciences, Kyushu University, Fukuoka, Japan
[3] Institute for Materials Science, University of Stuttgart, Stuttgart, Germany

Safe and high-density storage of hydrogen, for a clean-fuel economy, can be realized by hydride-forming materials, but these materials should be able to store hydrogen at room temperature. Some high-entropy alloys (HEAs) have recently been shown to reversibly store hydrogen at room temperature, but the design of HEAs with appropriate thermodynamics is still challenging. To explore HEAs with appropriate hydride formation enthalpy, this study employs machine learning (ML), in particular, Gaussian process regression (GPR) using four different kernels by training with 420 datum points collected from literature and curated here. The developed ML models are used to predict the formation enthalpy of hydrides for the $Ti_xZr_{2-x}CrMnFeNi$ ($x$ = 0.5, 1.0 and 1.5) system, which is not in the training set. The predicted values by ML are consistent with data from experiments and density functional theory (DFT). The present study thus introduces ML as a rapid and reliable approach for the design of HEAs with hydride formation enthalpies of -25 to -39 kJ/mol for hydrogen storage at room temperature.

***Keywords:*** solid-state hydrogen storage; multi-principal element alloys (MPEAs), high-entropy hydrides; artificial intelligence; ab initio calculations

*Corresponding author (E-mail: kaveh.edalati@kyudai.jp; Tel/Fax: +81 92 802 6744)

## 1. Introduction

With the increasing level of pollution in Earth's atmosphere, an alternative to fossil fuels is a necessity for society. Hydrogen is considered as an important alternative energy carrier. The energy carried by 1 kg of hydrogen is 142 MJ and this is greater than any other chemical fuel [1]. Storing hydrogen is an important challenge in the scientific and industrial communities because traditional storage methods such as gas storage and liquid storage have some drawbacks. For gaseous hydrogen, it requires a special tank design to store high-pressure hydrogen and these high-pressure tanks are associated with stringent safety codes and standards. For liquid hydrogen, the solidification temperature is 13.8 K whereas the vaporization temperature is 20.3 K and this small window to stabilize the liquid requires cryogenic applications and consumption of a significant amount of energy [1,2]. An alternative to these storage technologies is the use of metal hydrides which not only increase the volumetric capacity but also ensure safety due to low storage pressures [3,4].

A major issue associated with metal hydrides is the requirement of using high temperatures either for initial activation (kinetics issue) or for releasing hydrogen (thermodynamics issue) [3]. Although some alloys such as TiFe, Ti-V-Cr, $TiMn_2$ and $LaNi_5$ can thermodynamically release hydrogen at room temperature, only $LaNi_5$ shows a perfect performance in terms of kinetics, activation and reversibility [3,4]. However, $LaNi_5$ has a rather low capacity of 1.4 wt% and it contains rare and expensive elements. A few recent studies suggested that some high-entropy alloys (HEAs), i.e., alloys containing typically five principal elements, can exhibit fast and reversible hydrogen storage of up to 1.8 wt% at room temperature without a need for high-temperature activation treatment [5-7]. The good performance of HEAs at room temperature is intrinsically linked to the thermodynamic concept of the Gibbs energy ($\Delta G$),

$$\Delta G = \Delta H - T\Delta S, \qquad (1)$$

where $\Delta H$ and $\Delta S$ are the enthalpy and entropy of hydride formation and $T$ is the temperature. The major contribution to the entropy stems from hydrogen that changes from the gaseous to the solid phase, while the entropy change between metal and hydride has a minor contribution. The entropy of hydride formation generally lies between -110 to -130 J/mol·K for various hydrogen storage materials [8-11], and it is in the same range for HEAs as will be shown in this article. The enthalpy of hydride formation, which is experimentally measured using the van't Hoff analysis [1,2], is a central thermodynamic term that determines the performance of a material for hydrogen storage. Hence, the prediction of this parameter is a critical issue in designing HEAs for room-temperature hydrogen storage applications.

Although some descriptors were introduced for achieving the desired crystal structures, such as the C14 Laves phase in high-entropy hydrogen storage materials [5-7], the design of HEAs with appropriate thermodynamics still relies on experimental trials, mainly using pressure-composition-temperature (PCT) isotherm measurements. Density functional theory (DFT), which uses electron density to calculate the electronic structure without going through the complex calculation of the many-electron wave function [12], is one approach to determine the hydride formation enthalpy. DFT has proved to be useful for the design of solid-state hydrogen storage materials [13, 14]. For example, DFT has led to the design of the first Mg-based alloy with a hydrogen binding energy close to -0.1 eV and room-temperature hydrogen storage capability [15]. However, DFT calculations of HEAs are complicated and time-consuming due to the presence of numerous elements. A complementary computational method is the application of artificial intelligence via machine learning (ML), which is gaining popularity in various scientific fields due to its fast and reliable predictions (if used appropriately). ML facilitates the construction of complex relationships between input and output data without the need for explicit analytical relationships. ML has proved successful in various scientific fields such as medical [16,17], aerospace [18,19], energy [20], materials science [21-23], etc. ML algorithms have been employed in the hydrogen storage field for predicting storage capacity [24-26], hydride formation enthalpy [8,27,28], classification of metal hydrides [29] and exploring new metal hydrides by considering economic and technical feasibility [30,31]. However, the advantages of ML have not yet been exploited for the prediction of the enthalpy of hydride formation to design HEAs for room-temperature hydrogen storage. Perhaps one reason for the lack of such studies is the scarcity of available data and their spread over many different studies. A limited input dataset can easily make the output of ML unreliable.

In the present study, ML is employed to predict the enthalpy of hydride formation in HEAs by fitting to a dataset of 420 datum points collected from literature and curated here. The validity of the fitted ML models is examined by calculating the hydride formation enthalpy for $AB_2$-type (A: elements that form hydride; B: elements that do not form hydride) $Ti_xZr_{2-x}CrMnFeNi$ ($x$ = 0.5, 1.0, 1.5) alloys with room-temperature hydrogen storage capability [5,13]. It is found that the prediction of ML is consistent with experimental measurements obtained with the van't Hoff analysis and with first-principles calculations based on DFT. These findings introduce ML as a powerful and fast tool for designing high-entropy hydrogen storage materials.

## 2. Materials and Methods

### 2.1 Experimental Procedure

Ingots of $Ti_xZr_{2-x}CrMnFeNi$ ($x$ = 0.5, 1.0, 1.5) were prepared by arc melting of high-purity titanium (99.9%), zirconium (99.2%), chromium (99.99%), manganese (99.9%), iron (99.9%) and nickel (99.99%). The ingot was synthesized in the vacuum in a water-cooled copper mold, according to the procedure published earlier [6,7]. Rotating and remelting of the ingots were done seven times to achieve a homogeneous composition.

For structural analysis, a piece of the HEAs was crushed and examined using X-ray diffraction (XRD) with Cu Kα radiation, a filament current of 200 mA and an acceleration voltage of 45 kV. The analysis of XRD data was done using the PDXL software.

For the microstructural examination, cylindrical discs with a height of 0.8 mm and a diameter of 10 mm were prepared from the ingot using electrical discharge machining. Grinding and polishing were carried out on the discs using emery papers (grit: 400, 800, 1200 and 2000) and buff (9 µm diamond suspension, 3 µm diamond suspension and 60 nm colloidal silica) to obtain mirror-like surfaces. The microstructure of the mirror-like polished discs was examined using a scanning electron microscope (SEM) with an acceleration voltage of 15 kV. The discs were examined further for phase structure and chemical composition using electron backscatter diffraction (EBSD) and energy-dispersive X-ray spectroscopy (EDS).

The ability of the HEA to absorb-desorb hydrogen was tested using a Sievert-type machine. A piece of the HEAs was crushed in air and passed through a sieve with a mesh size of 75 µm. The obtained sample was kept under vacuum in the chamber of the Sievert-type equipment for 3 h at room temperature to remove possible humidity. Subsequently, hydrogen PCT isotherms were determined for 3 cycles at room temperature (303 K). To confirm the crystal structure of the hydride, the sample was hydrogenated again after the third PCT cycle and immediately examined using XRD. To measure the enthalpy and entropy of hydride formation through the van't Hoff analysis, another sample was inserted into the Sievert-type machine, and PCT isotherms were measured at different temperatures. Initially, one cycle was performed at room temperature to stabilize the activity, followed by several other cycles at temperatures of 303, 318, 327, 353, 363, and 423 K. For each PCT isotherm, the region where the plot of ln($P/P_0$) ($P$: hydrogen pressure, $P_0$ = 1 atm) versus hydrogen content could be fitted by a straight line was considered the plateau region. To accurately identify this region, the first derivative of the plots was also calculated, and the region with the minimum derivative was designated as the plateau region. After locating the plateau region, the pressure at the midpoint of this region was measured as the equilibrium plateau pressure ($P_{eq}$) and used for the van't Hoff analysis. It should be noted that in the van't Hoff analysis, it is assumed that hydrogen storage follows the following ideal form:

$$\mathrm{M} + \frac{x}{2}\mathrm{H_2} \rightleftharpoons \mathrm{MH}_x + \mathrm{Energy} \quad (2)$$

Therefore, it is possible to make a correlation between the equilibrium plateau pressure ($P_{eq}$) measured by the PCT isotherms and the thermodynamic parameters [1,2]:

$$\ln\frac{P_{\mathrm{eq}}}{P_0} = \frac{\Delta H}{RT} - \frac{\Delta S}{R} \quad (3)$$

In this equation, $P_0$ is the atmospheric pressure and $R$ is the gas constant. To obtain $\Delta H$ and $\Delta S$, plateau pressures should be measured at least at two temperatures.

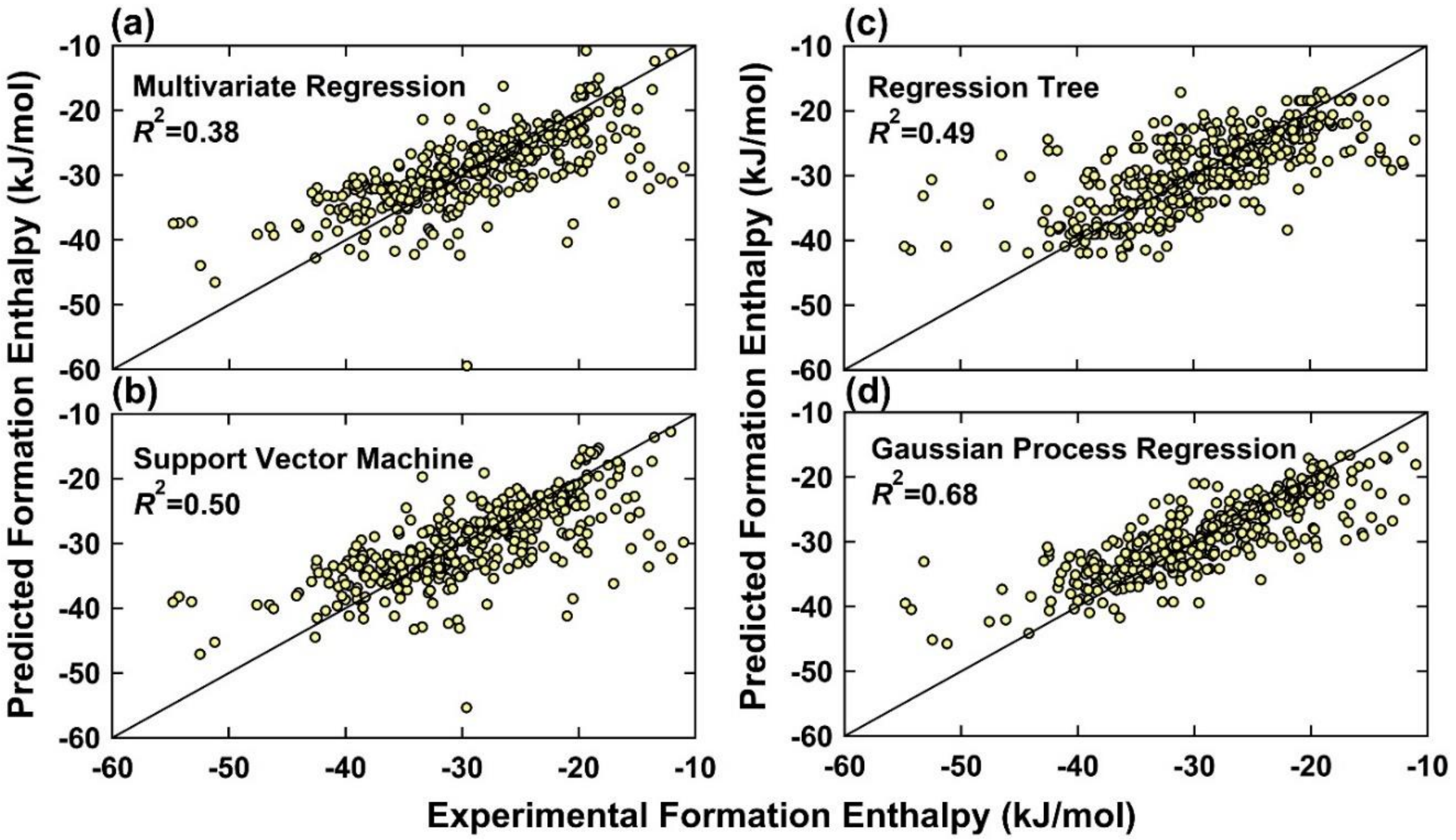

Fig. 1. Predicted formation enthalpy values plotted versus the measured experimental formation enthalpy using four different ML algorithms of (a) multivariate regression, (b) support vector machine, (c) regression tree, (d) Gaussian process regression (GPR). GPR showed the highest coefficient of determination ($R^2$) among the four ML algorithms.

**2.2 Machine Learning**

A dataset with 420 data was used for the ML modeling. The data were extracted from literature and correspond to $AB_2$-type materials used for hydrogen storage, as summarized in the Appendix. Objective factors, including elemental composition and existing hydride formation enthalpy for $AB_2$-type materials, were considered, while no subjective factors were included in the modeling. Since hydrogen storage properties are mainly affected by the composition of the alloys [13,32], the ML input parameters were the compositions of the relevant 24 elements in atomic percent (lithium, boron, carbon, magnesium, aluminum, silicon, titanium, vanadium, chromium, manganese, iron, cobalt, nickel, copper, yttrium, zirconium, niobium, molybdenum, tin, lanthanum, cerium, gadolinium, holmium and tungsten). Although no data with a similar elemental combination to HEAs Ti-Zr-Cr-Mn-Fe-Ni exist in the dataset, all elements in this alloying system exist in different combinations in the dataset. The output of the modeling was the hydride formation enthalpy. Since experimental values for the formation enthalpy can be slightly different depending on whether the hydrogen absorption or desorption plateau pressure is used, the enthalpy of desorption was taken for the present study, as suggested in an earlier study [2]. If the enthalpy was not given explicitly in a paper, it was calculated using the van't Hoff analysis by assuming that the entropy change equals to -110 J/mol·K [8]. To select the most appropriate ML model, the data were first trained using four different algorithms, including (a) multivariate regression, (b) support vector machine, (c) regression tree and (d) Gaussian process regression (GPR), as shown in Fig. 1. The algorithm with the highest *R*-squared ($R^2$) value, which was GPR, was selected for ML modeling.

Modeling of the 420 datum points from the literature was done by GPR, with a procedure as shown schematically in Fig. 2. GPR creates a prior distribution in the absence of data and then gradually transforms it to a posterior distribution with the datum points [33]. The similarity in datum points in ML is measured by a covariance matrix known as the kernel. The input datum points having similar values lie close to each other and their corresponding output must lie close to each other too [34,35]. Four kernels [20] were used in the present study as given in the following.

Exponential kernel:

$$k_{\text{Exponential}}(r) = \sigma^2 \exp\left(-\frac{||r||}{l}\right) \quad (4)$$

Rational quadratic kernel:

$$k_{\text{Rational Quadratic}}(r) = \sigma^2 \left(1 + \frac{||r||^2}{2\alpha l^2}\right)^{-\alpha} \quad (5)$$

Squared exponential kernel:

$$k_{\text{Squared Exponential}}(r) = \sigma^2 \exp\left(-\frac{||r||}{2l^2}\right) \quad (6)$$

Matern kernel:

$$k_{\text{Matern}}(r) = \sigma^2 \left(1 + \frac{\sqrt{5}||r||}{l} + \frac{5||r||^2}{3l^2}\right) \exp\left(-\frac{\sqrt{5}||r||}{l}\right) \quad (7)$$

In these kernels, $||r||$ is the norm Euclidean distance which is given by

$$||r|| = \sqrt{(c_i - c_j)^T (c_i - c_j)}, \quad (8)$$

where $c_i$ and $c_j$ are the compositions of the two different alloys in at%. For the $i^{th}$ alloy in a dataset of $n$ alloys, with a formation enthalpy of $\Delta H_i$, $c_i$ is a vector with components as $c_{i1}$, $c_{i2}$, ……,$c_{im}$, where $m$ represents the number of elements. Similarly, for the $n^{th}$ alloy, $c_n$ is a vector with components as $c_{n1}$, $c_{n2}$, ……,$c_{nm}$ ($c_{nm}$: composition of the $m^{th}$ element in the $n^{th}$ alloy). $\sigma^2$, $l$ and $\alpha$ are the signal variance, characteristic length scale and a parameter to determine the weight of large and small variations, respectively. The Matern kernel was used with a shape parameter value of 2.5 in the present study. The performance of each kernel was assessed by statistical factors defined as follows [32,34]:

$$\text{Coefficient of Determination: } R^2 = 1 - \frac{\sum_{i=1}^{n}\left(\Delta H_i^{\text{pred}} - \Delta H_i\right)^2}{\sum_{i=1}^{n}(\Delta H_i - \overline{\Delta H_i})^2} \quad (9)$$

$$\text{Root Mean Squared Error: (RMSE)} = \sqrt{\frac{\sum_{i=1}^{n}\left(\Delta H_i^{\text{pred}} - \Delta H_i\right)^2}{n}} \quad (10)$$

$$\text{Mean Squared Error: (MSE)} = \frac{1}{n}\sum_{i=1}^{n}\left(\Delta H_i^{\text{pred}} - \Delta H_i\right)^2 \quad (11)$$

$$\text{Mean Absolute Error: (MAE)} = \frac{1}{n}\sum_{i=1}^{n}|\Delta H_i^{\text{pred}} - \Delta H_i|, \quad (12)$$

where $\Delta H_i$ is the input enthalpy of the $i^{\text{th}}$ alloy and $\Delta H_i^{\text{pred}}$ is the predicted value from the GPR algorithm.

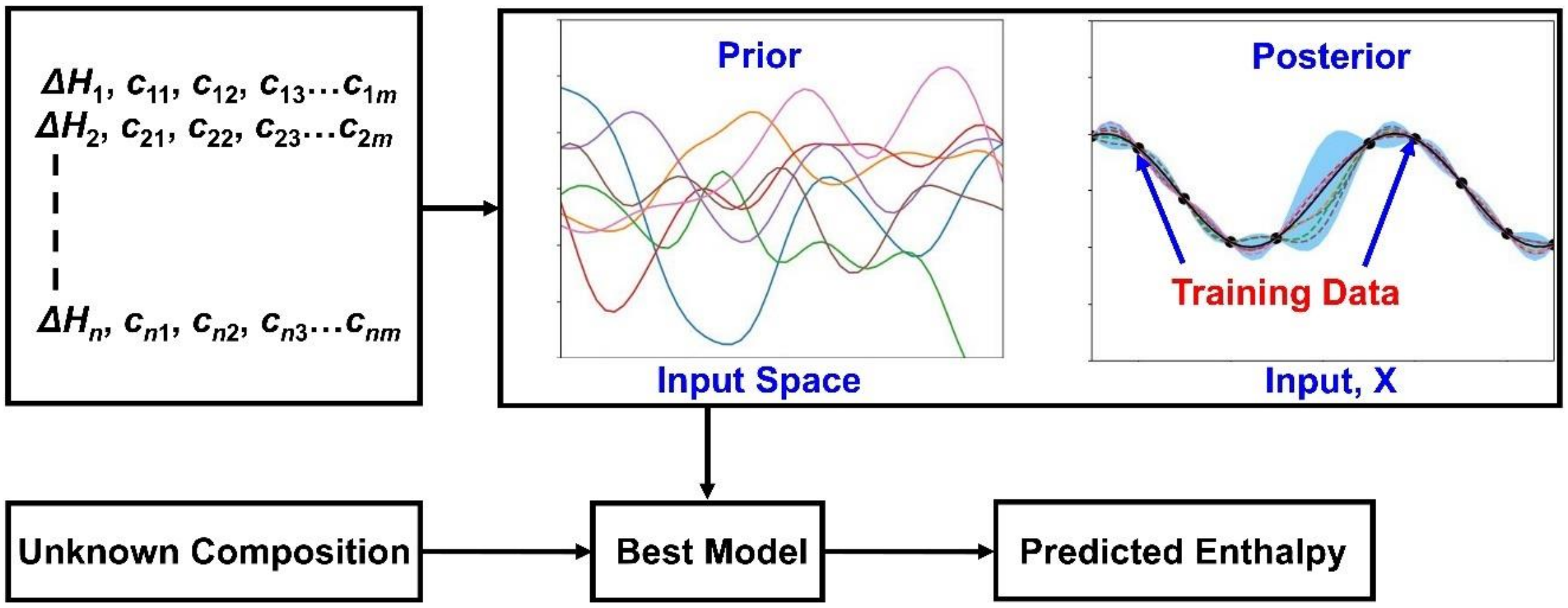

Fig. 2. Illustration of Gaussian process regression, in which the enthalpy ($\Delta H_n$) is the dependent variable and the composition ($c_{nm}$) is the independent variable of the $m^{\text{th}}$ element in the $n^{\text{th}}$ alloy. Incorporating the training data into the prior refines the posterior, giving a better relation between independent and dependent variables.

The enthalpy of each alloy was taken for training the ML GPR models utilizing the MATLAB software. Two types of methods were employed to train the data. The first method was *K*-fold cross-validation [23] with $K = 5$. The *K*-fold cross-validation was preferred over hold-out validation due to its ability to yield results with a higher accuracy range, typically ranging from 0.1% to 3% [36]. The second method was self-validation, in which all the data were used to train and test the ML model. The performance of both 5-fold cross-validation and self-validation was improved by curation, i.e., by removing the outliers from the data utilizing leverage analysis. The leverage analysis was conducted by calculating the hat matrix. The diagonal of the hat matrix was plotted against the standardized residuals and the values outside the designated sound zone and critical leverage value were considered outliers [37-39]:

$$\text{Hat Matrix} = C(C^T C)^{-1}C^T \quad (13)$$

$$\text{Critical Leverage} = \frac{3(m+1)}{n} \quad (14)$$

where $C$ is an $n \times m$ matrix which indicates the composition of the alloys with $n$ the number of datum points and $m$ the number of independent variables. The outliers were removed for the two methods by the leverage method with a critical leverage of 0.17. The inlier sound zone for the standardized residual was between -2 and 2 for 5-fold cross-validation and between -3 and 3 for self-validation. Williams plots for the outlier detection are shown in Fig. 3. The inliers identified in these plots were used to train the ML GPR models. The outliers of the dataset are marked with an asterisk in Table A1 of the Appendix.

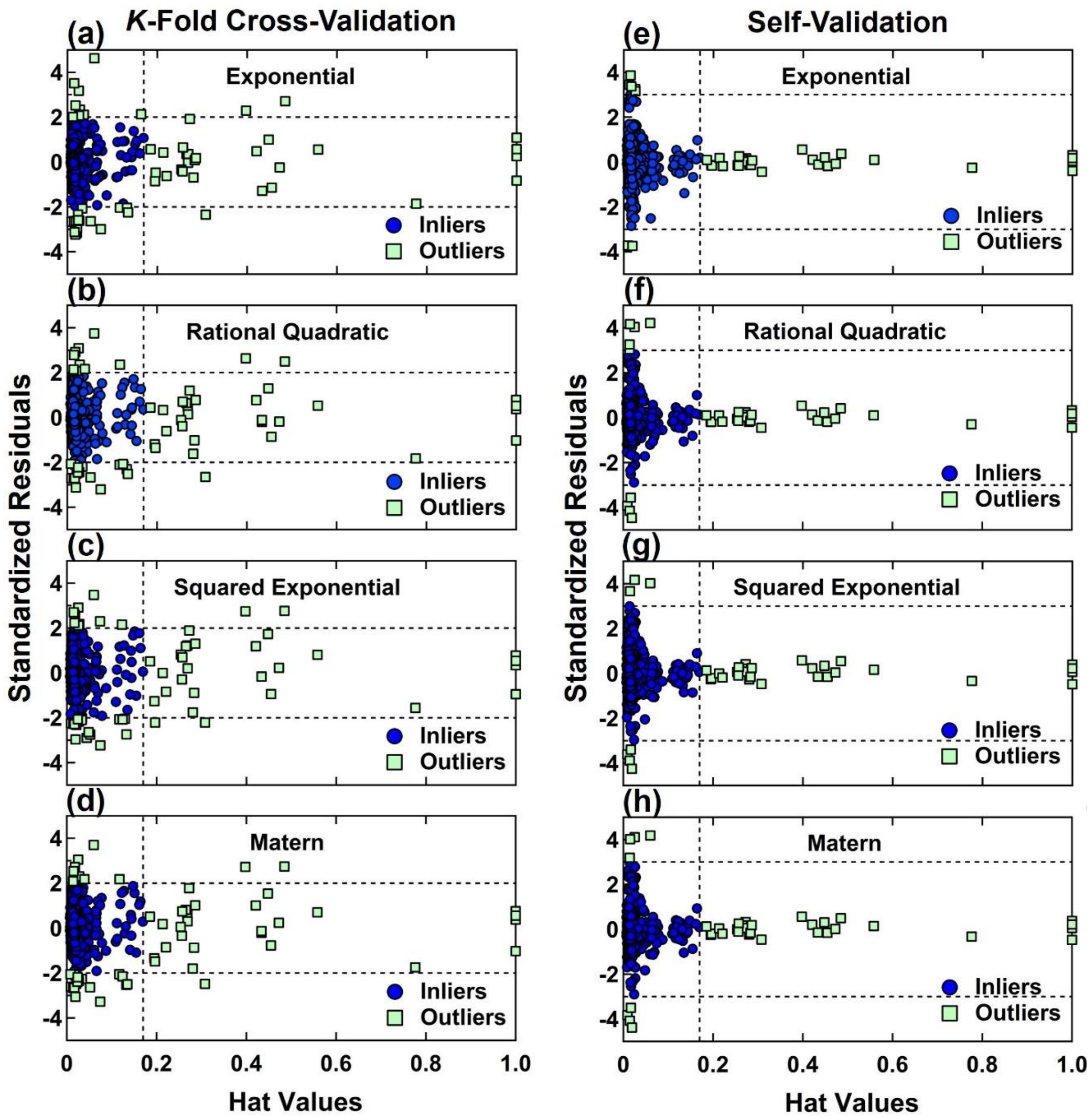

Fig. 3. Williams plots of standardized residuals versus hat values for (a, b, c, d) 5-fold cross-validation and (e, f, g, h) self-validation using four ML GPR kernels. The dotted lines represent the critical leverage and sound zone for standardized residuals (any datum points lying outside these dotted lines are considered outliers).

### 2.3 First-Principles Calculations

$Ti_xZr_{2-x}CrMnFeNi$ HEAs with the C14 Laves phase [5] were modeled based on a 48-atom supercell, for which the primitive C14 lattice was expanded by 2 × 2 × 1. Titanium and zirconium were placed in the A sites whereas chromium, manganese, iron and nickel were placed in the B sites [40]. Chemical disorder within the sublattice was simulated through the utilization of special quasi-random-structure (SQS) configurations [41]. The correlation functions for the initial nearest-neighbor clusters were fine-tuned through simulated annealing using the ICET code [42]. For better statistics, six different configurations were considered for each composition by permuting the elements. For hydride modeling, two assumptions were used: (i) Laves phase hydrides with the stoichiometry of $AB_2H_3$ are formed from Laves phase alloys with the stoichiometry of $AB_2$ [5], and (ii) hydrogen atoms exclusively occupy $A_2B_2$ sites characterized by the Wyckoff symbol 12k. The second assumption is motivated by considering that these sites avoid sharing faces with the interstitial tetrahedra, ensuring that the hydrogen atoms residing in them maintain a sufficient distance from each other [43]. Seven volumes were considered for HEAs and hydrides, and then the energies were fitted to the Vinet equation of state [44,45] to determine the equilibrium volumes. The atomic positions were relaxed and energetically optimized until all atomic forces were below 0.05 eV/Å. Two examples of optimized supercells for TiZrCrMnFeNi and $TiZrCrMnFeNiH_6$, visualized using OVITO [46], are shown in Fig. 4.

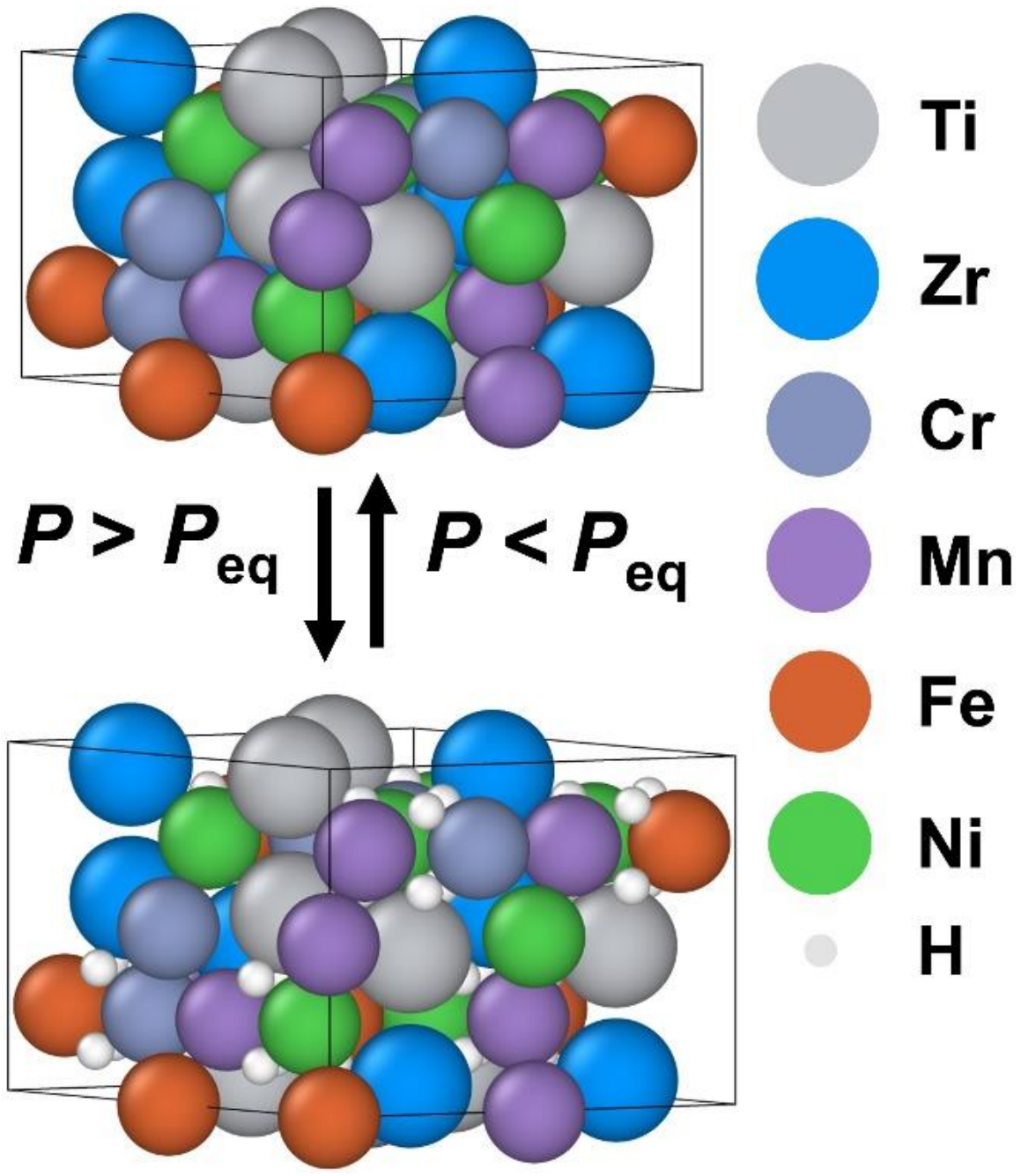

Fig. 4. Optimized supercells for first-principles DFT calculations of TiZrCrMnFeNi and $TiZrCrMnFeNiH_6$. The figure illustrates the mechanism of hydrogen storage by hydrogen absorption in a high-entropy alloy and hydrogen desorption from a high-entropy hydride (*P*: hydrogen pressure, $P_{eq}$: equilibrium plateau pressure).

DFT calculations were performed using the Vienna ab initio simulation package (VASP) [47-49]. Electron-ion interactions were modeled using the projector augmented wave (PAW) method and a plane-wave basis set [50] with a cutoff energy of 400 eV. The exchange-correlation energy was computed employing the Perdew-Burke-Ernzerhof (PBE) form of the generalized gradient approximation (GGA) [51]. A Γ-centered 4 × 4 × 6 *k*-point mesh was chosen to sample the reciprocal space of the 48-atom supercell in combination with the Methfessel-Paxton method [52] with a smearing width of 0.1 eV. Total energies were iteratively minimized until convergence was achieved within $1 \times 10^{-3}$ eV per simulation cell for each ionic step, with all computations incorporating spin polarization. The formation enthalpy of hydride at 0 K was calculated using the following relationship by considering the enthalpies of HEAs, high-entropy hydrides and hydrogen molecules. Although the dependence of enthalpy on temperature was not calculated due to the computational time limits, the formation enthalpies at 0 K can be reasonably used at room temperature, provided that the heat capacities of alloys and hydrides are assumed to be similar.

$$\Delta H = H(\mathrm{AB_2H_3}) - H(\mathrm{AB_2}) - \frac{3}{2}H(\mathrm{H_2}) \quad (15)$$

where $H(\mathrm{AB_2H_3})$, $H(\mathrm{AB_2})$ and $H(\mathrm{H_2})$ are the enthalpies per mole of hydride, HEA and a hydrogen molecule, respectively. The energy of the hydrogen molecule was computed in a 20 × 20 × 20 $Å^3$ simulation cell and for the Γ point in the reciprocal space. The optimized hydrogen-hydrogen distance was 0.751 Å, in good agreement with the experimental value of 0.74144 Å [53].

# 3. Results

## 3.1 Experimental Results

The PCT isotherms at 303 K are shown in Fig. 5 for the $Ti_xZr_{2-x}CrMnFeNi$ HEAs with (a) $x$ = 0.5, (b) $x$ = 1.0 and (c) $x$ = 1.5. $Ti_{0.5}Zr_{1.5}CrMnFeNi$ absorbs 1.6 wt% hydrogen at room temperature reversibly without any high-temperature activation. The first PCT cycle shows a high plateau pressure of around 0.15 MPa with a high hysteresis between the adsorption and desorption plateau pressures. The hysteresis becomes negligible and absorption plateau pressure decreases to ~0.1 MPa for the second and third cycles. Despite these pressure alternations, the storage capacities are reasonably the same for the three cycles. The higher absorption pressure and hysteresis in the first cycle of hydrogen storage are due to the initial energy required for the activation of the metallic phase to uptake hydrogen [2]. Despite high pressure in the first cycle due to the activation and kinetics issues [2], the plateau pressure in the second and third cycles is reduced and mainly affected by thermodynamic issues such as the enthalpy and entropy of hydride formation. For equiatomic-composition HEA TiZrCrMnFeNi, the material also reversibly absorbs 1.6 wt% hydrogen at room temperature without any high-temperature activation treatment. The first three hydrogenation and dehydrogenation cycles show similar behavior with minor hysteresis and similar storage capacities, indicating good cyclic capability and high activity of this alloy even in the first PCT cycle. $Ti_{0.5}Zr_{1.5}CrMnFeNi$ absorbs a small amount of ~0.1 wt% hydrogen at room temperature without showing any plateau pressure, indicating that the plateau pressure for this material is higher than 9 MPa which is the maximum achievable pressure in the authors' gas absorption facility. The hydrogen storage amount of 1.6 wt% obtained for the two alloys $Ti_{0.5}Zr_{1.5}CrMnFeNi$ and TiZrCrMnFeNi is considered a good capacity for room-temperature hydrogen storage materials, particularly because the current materials do not need any extra activation treatment [1-4].

The XRD profiles before and after hydrogenation are shown in Fig. 5 for HEAs $Ti_xZr_{2-x}CrMnFeNi$ with (d) $x$ = 0.5, (e) $x$ = 1.0 and (f) $x$ = 1.5. The three HEAs before hydrogenation have a C14 Laves phase structure with lattice parameters given in Table 1. The formation of the C14 Laves phase should be one reason for the good hydrogen storage properties of this alloying system because C14 is known to have a high potential for hydrogen storage [6,7,13,40]. After hydrogenation, new Laves phases of hydrides with the lattice parameters given in Table 1 are formed for $Ti_{0.5}Zr_{1.5}CrMnFeNi$ and TiZrCrMnFeNi, while $Ti_{1.5}Zr_{0.5}CrMnFeNi$ does not show the formation of Laves phase hydride due to its low stored hydrogen. The Laves phase hydrides show an 18% lattice expansion compared to the initial metallic Laves phases. This lattice expansion is similar to binary and ternary hydrides reported in the literature [54]. The presence of the initial metallic Laves phase after hydrogenation should be due to the fast desorption of hydrogen before the XRD test because the dehydrogenation of the Laves phase can be quite fast even at room temperature [55-57].

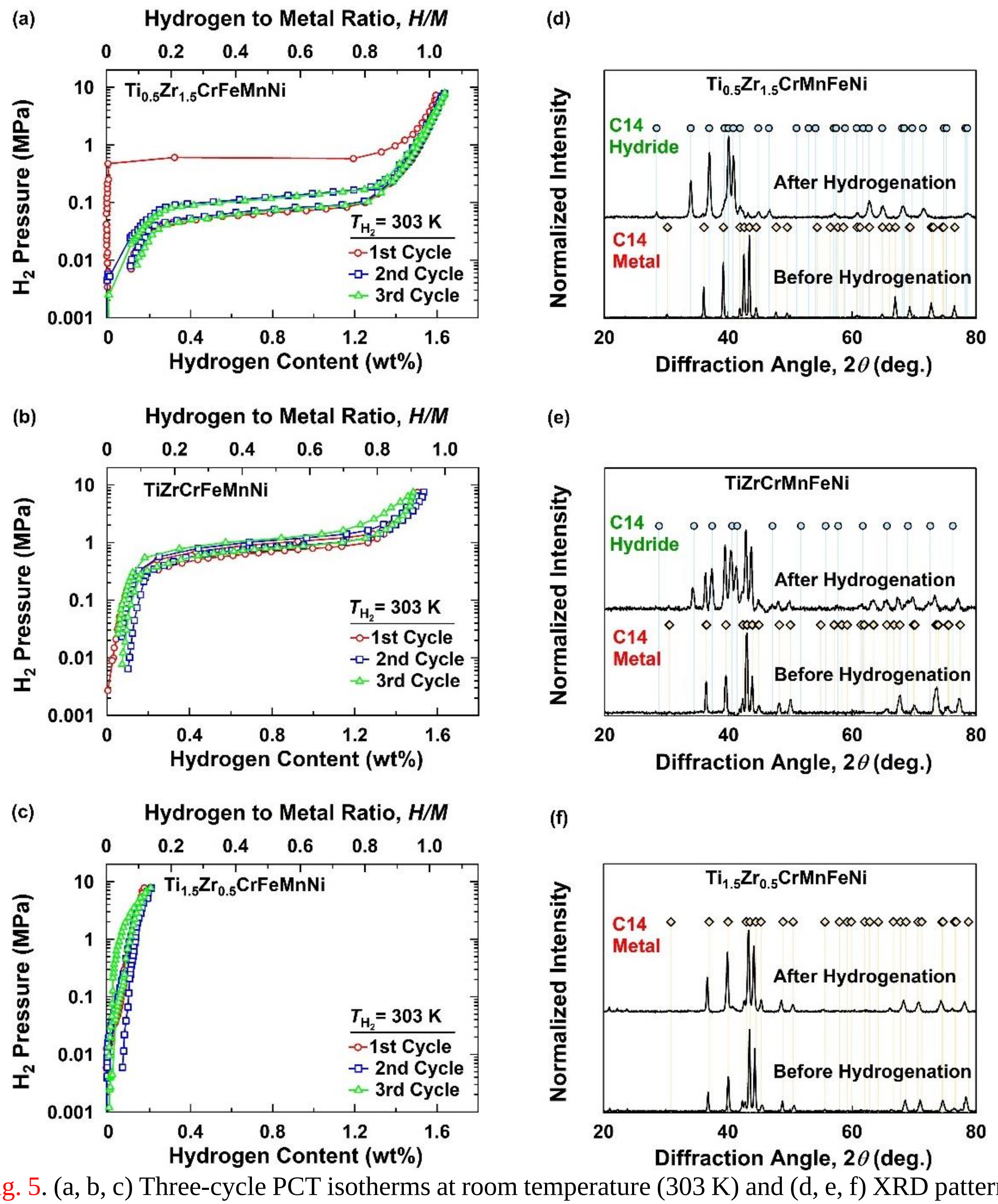

Fig. 5. (a, b, c) Three-cycle PCT isotherms at room temperature (303 K) and (d, e, f) XRD patterns before and after hydrogenation for HEAs (a, d) $Ti_{0.5}Zr_{1.5}CrMnFeNi$, (b, e) TiZrCrMnFeNi and (c, f) $Ti_{1.5}Zr_{0.5}CrMnFeNi$. The main phase is the C14 Laves phase before hydrogenation, and it transforms to another C14 laves phase after hydrogenation of $Ti_{0.5}Zr_{1.5}CrMnFeNi$ and TiZrCrMnFeNi, while $Ti_{1.5}Zr_{0.5}CrMnFeNi$ does not store an appreciable amount of hydrogen.

The examination of the microstructure of TiZrCrMnFeNi by SEM is shown in Fig. 6(a) and the corresponding EBSD phase mapping is in Fig. 6(b). EBSD confirms the presence of the C14 Laves phase (red color), but a ~2 vol% of BCC phase (blue color) is also detected. The BCC phase could not be detected in XRD because of its small fraction and the overlap of its peaks with the C14 phase. Fig. 6(c) shows EDS elemental mappings, confirming that the BCC phase is rich in titanium and nickel. Although the fraction of BCC is too small to affect the thermodynamics of hydrogen storage, it can be beneficial for easy activation and fast kinetics [6,7,58]. A quantitative analysis of the compositions of $Ti_{0.5}Zr_{1.5}CrMnFeNi$, TiZrCrMnFeN and $Ti_{1.5}Zr_{0.5}CrMnFeNi$ using SEM-EDS is given in Table 1, indicating a good agreement between the experimentally measured and nominal compositions.

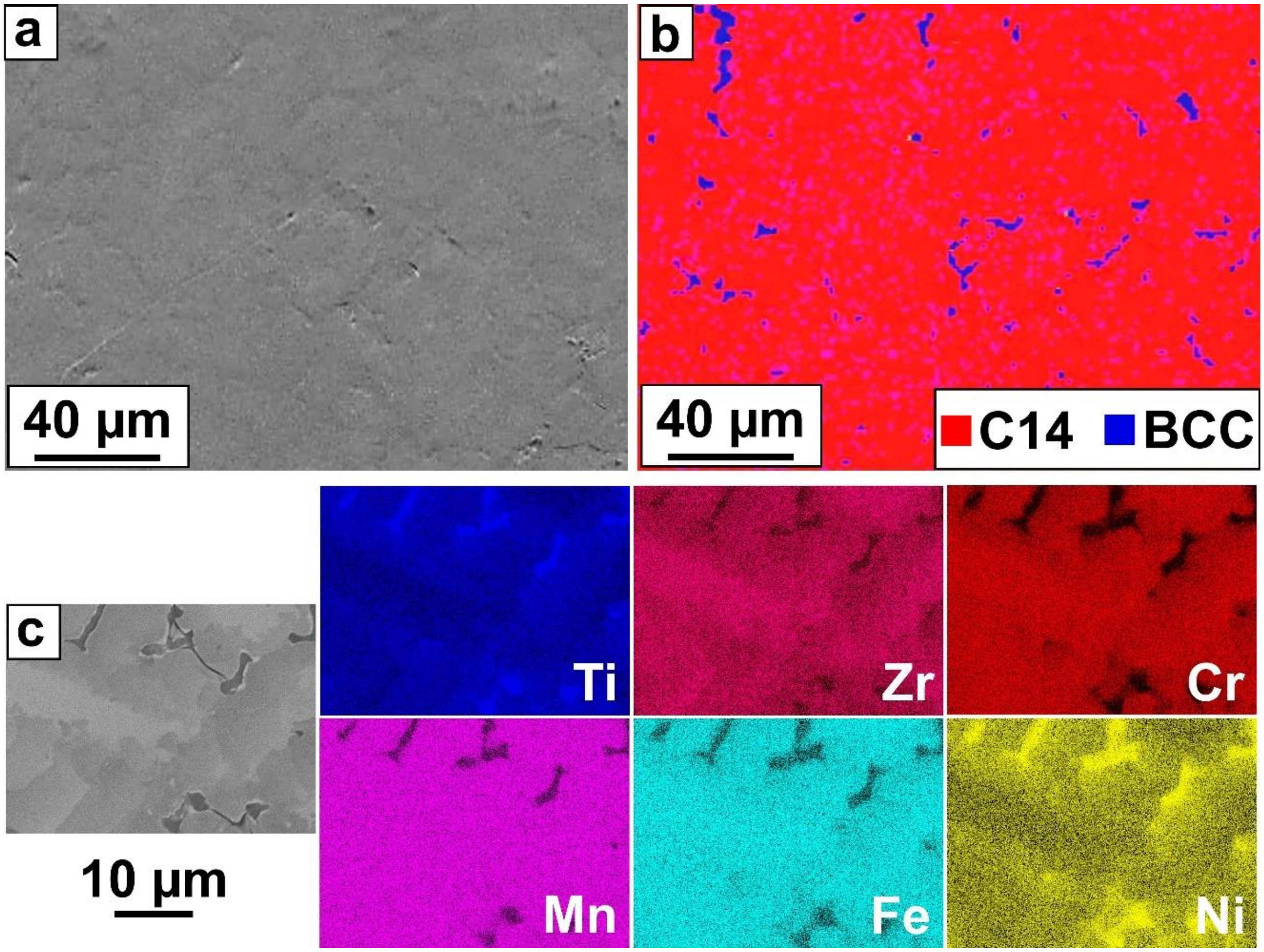

Fig. 6. (a) SEM micrograph and (b) corresponding EBSD phase map with a beam step size of 1 µm, and (c) SEM micrograph and corresponding EDS elemental mappings for the TiZrCrMnFeNi HEA. In addition to the C14 Laves phase, a small amount of a Ti- and Ni-rich BCC phase is also observed.

Table 1. Lattice Parameters and chemical compositions of $Ti_xZr_{2-x}CrMnFeNi$ HEA system ($x$ = 0.5, 1.0, 1.5) obtained through XRD and SEM-EDS.

| | **$Ti_{0.5}Zr_{1.5}CrMnFeNi$** | | **TiZrCrMnFeNi** | | **$Ti_{1.5}Zr_{0.5}CrMnFeNi$** | |
|---|---|---|---|---|---|---|
| **Lattice Parameters Before Hydrogenation** | $a = b$ = 0.498 nm, $c$ = 0.814 nm | | $a = b$ = 0.493 nm, $c$ = 0.807 nm | | $a = b$ = 0.490 nm, $c$ = 0.800 nm | |
| **Lattice Parameters After Hydrogenation** | $a = b$ = 0.527 nm, $c$ = 0.861 nm | | $a = b$ = 0.521 nm, $c$ = 0.854 nm | | No Hydride Phase | |
| **Compositions** | **Nominal** | **Measured** | **Nominal** | **Measured** | **Nominal** | **Measured** |
| **Ti (at%)** | 8.3 | 8.4 | 16.7 | 16.8 | 25.0 | 24.7 |
| **Zr (at%)** | 25.0 | 27.2 | 16.7 | 18.4 | 8.3 | 9.0 |
| **Cr (at%)** | 16.7 | 17.5 | 16.7 | 17.6 | 16.7 | 17.4 |
| **Mn (at%)** | 16.7 | 13.5 | 16.7 | 13.9 | 16.7 | 15.6 |
| **Fe (at%)** | 16.7 | 16.9 | 16.7 | 17.4 | 16.7 | 17.3 |
| **Ni (at%)** | 16.7 | 16.5 | 16.7 | 15.9 | 16.7 | 16.0 |

To measure the thermodynamic parameters relevant to hydrogen storage, the PCT desorption isotherms were obtained at different temperatures, as shown in Fig. 7 for (a) $Ti_{0.5}Zr_{1.5}CrMnFeNi$ and (b) TiZrCrMnFeNi. It should be noted that since $Ti_{1.5}Zr_{0.5}CrMnFeNi$ does not show a plateau pressure by increasing the hydrogen pressure to 9 MPa, it is not possible to evaluate its thermodynamic parameters using the authors' Sievert-type gas absorption machine. There is an increase in plateau pressure in Fig. 7 with increasing temperature, as expected from Eq. 3. The plateau pressure at 423 K and 363 K cannot be observed for TiZrCrMnFeNi because the hydrogenation/dehydrogenation pressures exceed the maximally achievable pressure of the gas absorption machine. The van't Hoff plot is shown in Fig. 7 obtained from datum points for (c) $Ti_{0.5}Zr_{1.5}CrMnFeNi$ and (d) TiZrCrMnFeNi at different temperatures. Evaluation of the slope and Y-intercept of the fitted line with the van't Hoff equation (Eq. 3) leads to a hydride formation enthalpy of $\Delta H$ = -36.5 kJ/mol and hydride formation entropy of $\Delta S$ = -117.1 J/mol·K for $Ti_{0.5}Zr_{1.5}CrMnFeNi$, and $\Delta H$ = -28.8 kJ/mol and $\Delta S$ = -110.4 J/mol·K for TiZrCrMnFeNi. The entropy values are similar to various binary and ternary hydrides (114.7 J/mol·K for $Ti_{1.1}CrMn$ [9], -122 J/mol·K for $ZrFe_2$ [10] and -116 J/mol·K for $TiCr_{1.8}$) [11]), indicating that entropy has no clear effect on the hydrogen storage behavior of the current HEAs. The examination of data reported in the literature for other HEAs also does not suggest any significant difference in entropy compared to conventional hydrogen storage materials (-108 to -122 J/mol·K for Al-Ti-V-Zr-Cr-Mn-Fe alloys [59] and -105 to -124 J/mol·K for Al-Ti-V-Zr-Cr-Mn-Fe-Co-Ni-Sn alloys [60]). Therefore, the formation enthalpy is responsible for reversible hydrogen storage in HEAs at room temperature. The target enthalpy for room-temperature hydrogen storage is usually in the range of -25 kJ/mol to -35 kJ/mol [61]. This range can be slightly shifted by the entropy effect. For example, to achieve reversible hydrogen storage at atmospheric pressure and room temperature, the logarithm of the pressure term in the van't Hoff equation (Eq. 1) becomes mathematically zero. Assuming the change in entropy lies between -110 to -130 J/mol for hydride formation [8-11], then the hydride formation enthalpy lies between -33 to -39 kJ/mol.

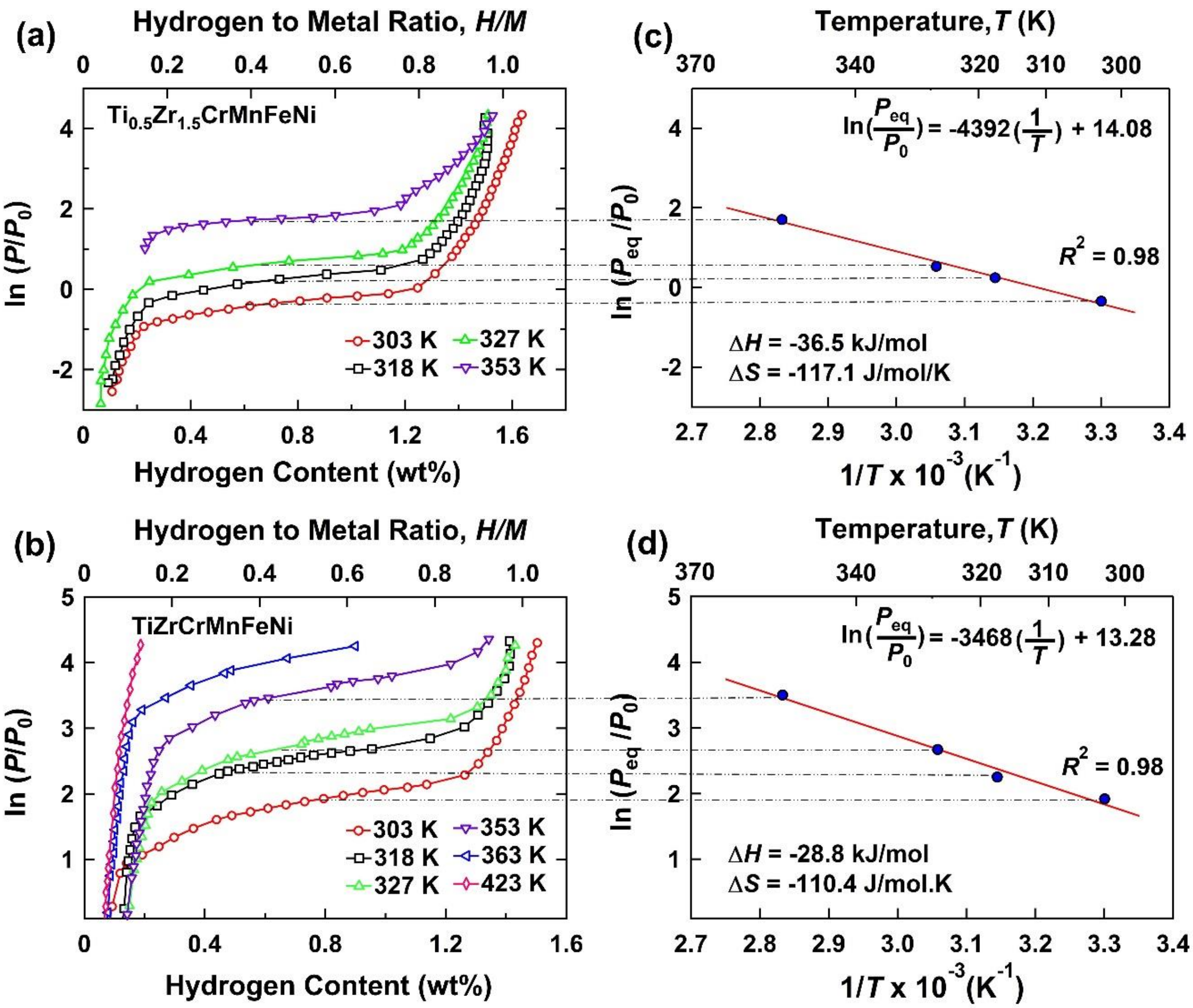

Fig. 7. (a, b) PCT desorption isotherms at different temperatures and (c, d) corresponding van't Hoff plot for HEAs (a, c) $Ti_{0.5}Zr_{1.5}CrMnFeNi$ and (b, d) TiZrCrMnFeNi ($P$: hydrogen pressure, $P_0$ = 1 atm, $P_{eq}$: equilibrium plateau pressure, $T$: temperature). Fitting of the van't Hoff plot has a high $R$-squared value of 0.98, and the calculated formation enthalpy and entropy of the hydride are $\Delta H$ = -36.5 kJ/mol and $\Delta S$ = -117.1 J/mol·K for $Ti_{0.5}Zr_{1.5}CrMnFeNi$, and $\Delta H$ = -28.8 kJ/mol and $\Delta S$ = -110.4 J/mol·K for TiZrCrMnFeNi.

### 3.2 Machine Learning Results

The model was first evaluated with the 5-fold cross-validation method. In this validation, the data were randomly split into five groups. The first group was used as a test set whereas the remaining 4 groups were used as training set. In the next step, the second group was changed to the test set and the remaining four groups were the training set. This procedure was repeated by changing the test set to the other groups to achieve different models. The final accuracy corresponds to the average accuracy of all five models. The second GPR method employed was self-validation where all the data was used to train the GPR models. The predicted value versus measured experimental value curve (i.e. cross-plot curve) for the 5-fold cross-validation method is shown in Fig. 8 for the (a) exponential, (b) rational quadratic, (c) squared exponential and (d)

Matern kernels. The $R^2$ values are between 0.64 to 0.68 with exponential GPR having the highest $R^2$. Fig. 8 also represents the measured experimental values versus predicted values for the GPR self-validation method for the (e) exponential, (f) rational quadratic, (g) squared exponential and (h) Matern kernels. The $R^2$ values are between 0.88 to 0.94 thus higher than those for 5-fold cross-validation. Similar to the 5-fold cross-validation method, the exponential kernel shows the highest $R^2$ for the self-validation method. Based on these ML models, the hydride formation enthalpy for HEAs $Ti_{0.5}Zr_{1.5}CrMnFeNi$ and TiZrCrMnFeNi, which were not a part of the fitting dataset, was predicted by each model. As shown in Fig. 8, all models predict the values reasonably. The closest value to the experimental value of -36.5 kJ/mol for $Ti_{0.5}Zr_{1.5}CrMnFeNi$ is predicted by the rational quadratic kernel in both 5-fold cross-validation and self-validation (-32.12 kJ/mol). The closest predicted formation enthalpy to the experimental value (-28.8 kJ/mol) of equiatomic HEA TiZrCrMnFeNi is predicted by the exponential kernel in both the 5-fold cross-validation and self-validation (-27.7 kJ/mol). For the HEA $Ti_{1.5}Zr_{0.5}CrMnFeNi$, the predicted formation enthalpy value is around -22 kJ/mol which is not in the range for room-temperature hydrogen storage materials [61] and hence it does not absorb hydrogen at room-temperature as shown by experiments in Fig. 5(c). These results together with earlier publications [24-31] confirm the high potential of ML in the hydrogen storage field.

The ML data after removing the outliers with the leverage method are shown in Fig. 9 for (a-d) 5-fold cross-validation and (e-h) self-validation using the four different kernels. The $R^2$ values for 5-fold cross-validation range from 0.67 to 0.75 which is better than without removing the outliers (0.64-0.68 in Fig. 8). The $R^2$ values for the self-validation method range from 0.93 to 0.98, which is quite high and better than without removing the outliers (0.88-0.94 in Fig. 8). For both validation methods, the exponential GPR models show the highest $R^2$. The closest predicted formation enthalpy for $Ti_{0.5}Zr_{1.5}CrMnFeNi$ was -31.7 kJ/mol by exponential GPR. The predicted enthalpy value for the TiZrCrMnFeNi using the exponential GPR and self-validation method is -27.3 kJ/mol which is only 5% different from the experimental value of -28.8 kJ/mol. Other statistical factors such as the root mean squared error (RMSE), mean squared error (MSE) and mean absolute error (MAE), as summarized in Table 2, also show that the self-validation method using the exponential kernel and after removing outliers exhibits the lowest values of RMSE, MSE and MAE, indicating high accuracy of the model [32,34]. These results suggest a strategy for the design of new materials for room-temperature hydrogen storage by ML, an artificial intelligence approach that has received attention in different fields [16-23], including hydrogen storage [24-31].

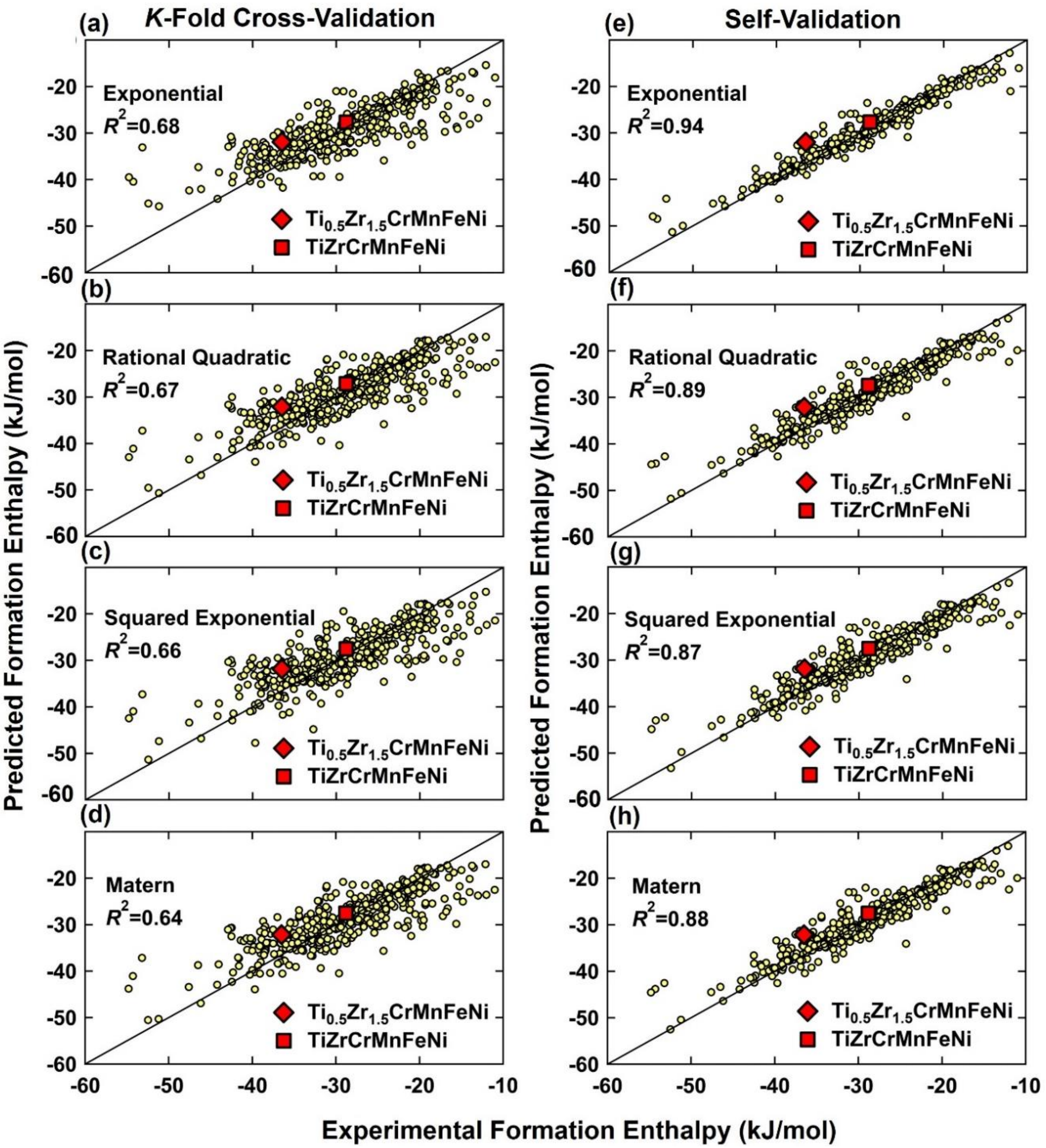

Fig. 8. Predicted formation enthalpy values plotted versus the measured experimental enthalpy values achieved for 420 datum points using (a-d) 5-fold cross-validation and (e-h) self-validation ML GPR methods with (a) exponential, (b) rational quadratic, (c) squared exponential and (d) Matern kernels. The diamond and square markers indicate the position of $Ti_{0.5}Zr_{1.5}CrMnFeNi$ and TiZrCrMnFeNi HEAs, respectively. The *R*-squared values are higher for the self-validation method and the exponential kernel provides the most accurate model.

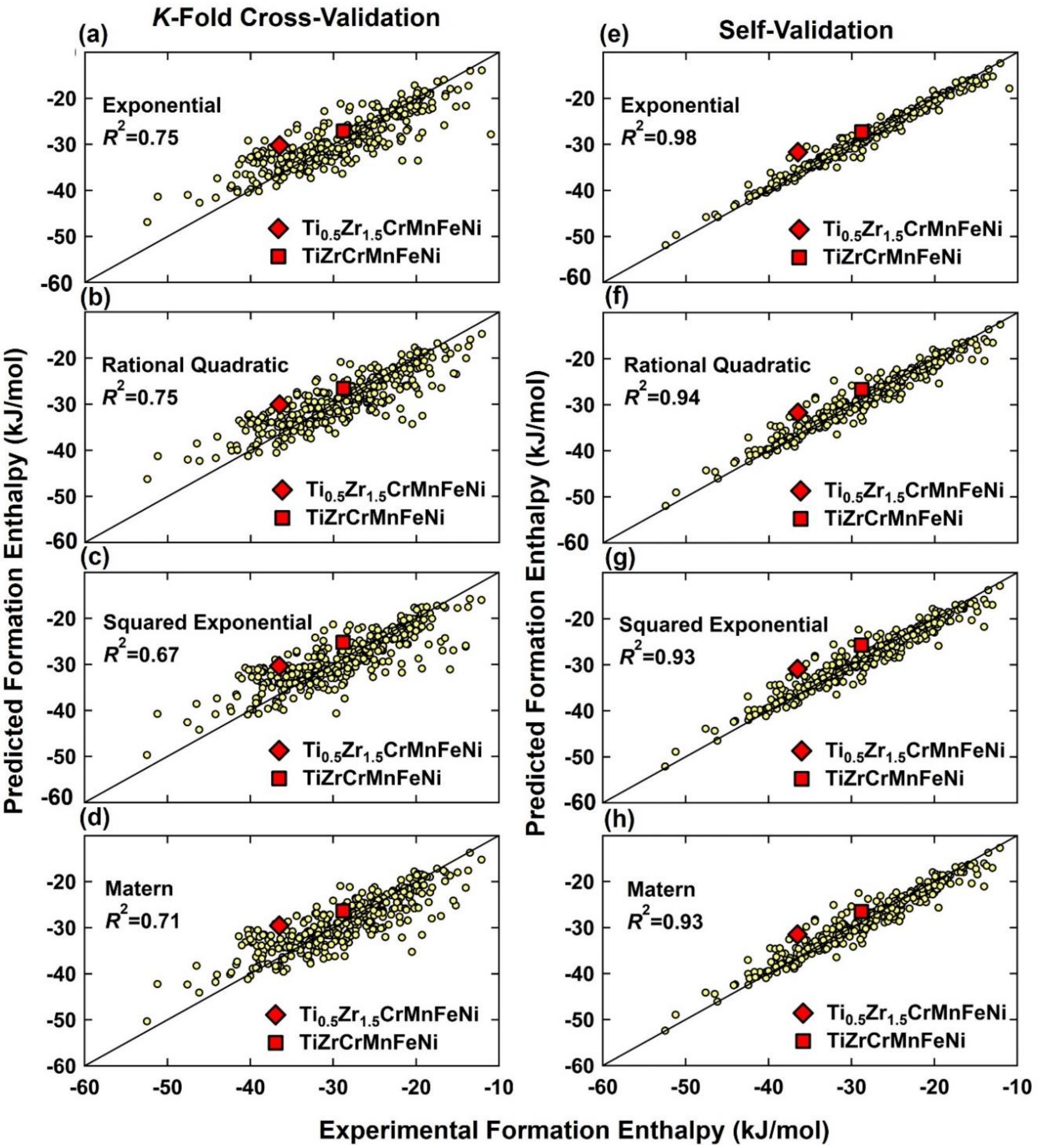

Fig. 9. Predicted formation enthalpy values plotted versus the measured experimental enthalpies achieved after removing outliers using (a-d) 5-fold cross-validation and (e-h) self-validation ML GPR methods with (a) exponential, (b) rational quadratic, (c) squared exponential and (d) Matern kernels. The diamond and square markers indicate the position of $Ti_{0.5}Zr_{1.5}CrMnFeNi$ and TiZrCrMnFeNi HEAs respectively. The *R*-squared values increase after removing outliers.

Table 2. Values of the statistical factors *R*-squared ($R^2$), root mean squared error (RMSE), mean squared error (MSE) and mean absolute error (MAE) for *K*-fold cross-validation and self-validation ML GPR methods without and after removal of outliers using different kernels.

| Kernel | *K*-Fold Cross-Validation | | | | Self-Validation | | | |
|---|---|---|---|---|---|---|---|---|
| | $R^2$ | RMSE | MSE | MAE | $R^2$ | RMSE | MSE | MAE |
| | Without Removing Outliers | | | | Without Removing Outliers | | | |
| **Exponential** | 0.68 | 4.22 | 17.79 | 3.01 | 0.94 | 1.77 | 3.14 | 1.14 |
| **Rational Quadratic** | 0.67 | 4.46 | 19.90 | 3.17 | 0.89 | 2.49 | 6.21 | 1.65 |
| **Squared Exponential** | 0.66 | 4.50 | 20.28 | 3.28 | 0.87 | 2.71 | 7.34 | 1.85 |
| **Matern** | 0.64 | 4.47 | 19.95 | 3.22 | 0.88 | 2.55 | 6.50 | 1.70 |
| | After Removing Outliers | | | | After Removing Outliers | | | |
| **Exponential** | 0.75 | 3.51 | 12.53 | 2.50 | 0.98 | 1.09 | 1.18 | 0.70 |
| **Rational Quadratic** | 0.67 | 3.62 | 13.12 | 2.53 | 0.94 | 1.76 | 3.08 | 1.20 |
| **Squared Exponential** | 0.75 | 3.68 | 13.57 | 2.69 | 0.93 | 1.91 | 3.64 | 1.35 |
| **Matern** | 0.71 | 3.87 | 14.95 | 2.68 | 0.93 | 1.84 | 3.38 | 1.28 |

### 3.3 Density Functional Theory Results

The hydride formation enthalpy using DFT was calculated for HEAs $Ti_xZr_{2-x}CrMnFeNi$ ($x$=0.5, 1.0, 1.5). After optimization of the supercell models, the calculated lattice expansion after hydrogenation was 21-23%, which is in reasonable agreement with the experimental value of 18% in this study. The hydride formation enthalpy was calculated as -36.5±0.61, -30.5±0.77 and -21.3±0.54 kJ/mol for $Ti_{0.5}Zr_{1.5}CrMnFeNi$, TiZrCrMnFeNi and $Ti_{1.5}Zr_{0.5}CrMnFeNi$, respectively. The calculated values of -36.5kJ/mol and -28.8 kJ/mol, which are in the range of enthalpies appropriate for room-temperature hydrogen storage [61], are so close to experimental formation enthalpies of -36.5 and -30.5 kJ/mol for $Ti_{0.5}Zr_{1.5}CrMnFeNi$ and TiZrCrMnFeNi, respectively. Moreover, the predicted formation enthalpy for $Ti_{1.5}Zr_{0.5}CrMnFeNi$ is not in the range of room-temperature hydrogen storage [61], and it is evident from the experiments that it does not absorb hydrogen. These results together with earlier publications [13-15] confirm the applicability of DFT in understanding the thermodynamics of hydrogen storage materials. Future DFT studies can be conducted to clarify the effect of lattice distortion on the mechanism underlying the good kinetics and activation of these inherently distorted HEAs because earlier experimental studies showed that materials distorted by external plastic strain can usually exhibit enhanced kinetics and good activation [62-64].

## 4. Discussion

This study successfully introduces ML for the design of HEAs for room-temperature hydrogen storage based on modeling of the hydride formation enthalpy. Here, three questions need to be discussed. First, how do the results of ML compare with the experimental and DFT values? Second, is it reasonable to use ML only for the prediction of enthalpy and not entropy? Third, which ML method gives the best accuracy?

Regarding the first question, Fig. 10 compares the hydride formation enthalpy for the $Ti_xZr_{2-x}CrMnFeNi$ HEAs ($x$ = 0.5, 1.0, 1.5) using experiments, ML via *K*-fold cross-validation and self-validation and DFT. A similar trend is observed in experiments, DFT and ML methods, i.e., the enthalpy becomes more negative with a decrease in the amount of titanium (i.e. the hydride becomes more stable). Although the ML models slightly overestimate the enthalpy and DFT slightly underestimates the enthalpy, the small differences are within the range of experimental errors either in this study or in the dataset gathered for the ML GPR models. Moreover, although DFT has high efficacy in electronic structure investigations [41-52] and enthalpy calculations of hydrogen storage materials [13-15]; it often underestimates the enthalpy for alloys containing elements with magnetic 3d orbitals when using GGA-PBE [51,65]. The calculated values for both ML and DFT are consistent with the experimental values achieved for $Ti_{0.5}Zr_{1.5}$CrMnFeNi and TiZrCrMnFeNi. Therefore, ML can be efficiently used for quick and reliable prediction of the enthalpy which is an important quantity for the design of hydrogen storage materials [1-4].

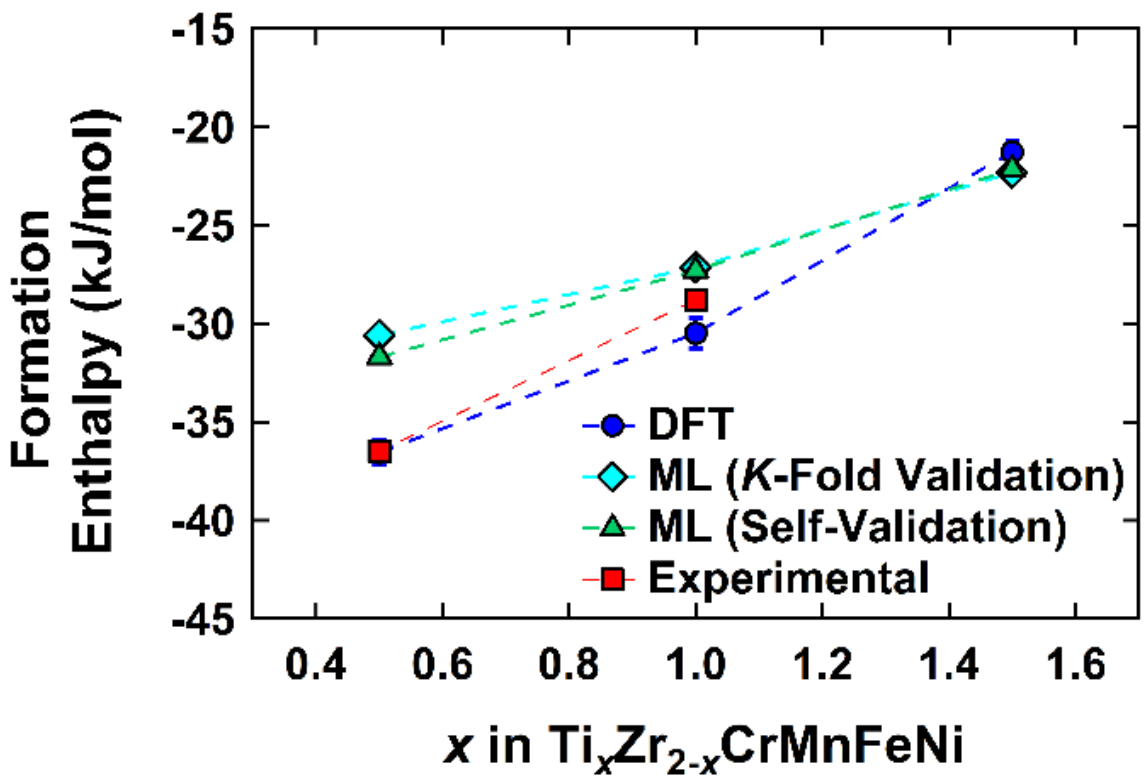

Fig. 10. Comparison of the formation enthalpy of hydride for HEAs $Ti_xZr_{2-x}$CrMnFeNi ($x$ = 0.5, 1.0, 1.5) obtained by experiments via the van't Hoff analysis, first-principles calculations via DFT, ML GPR via 5-fold cross-validation and self-validation using the exponential kernel and after removing outliers. The formation enthalpy becomes less negative with increasing titanium and there is good consistency between DFT, ML and experiments.

Regarding the second question, the present study shows that the entropy of $Ti_{0.5}Zr_{1.5}$CrMnFeNi is -117.1 J/mol·K and TiZrCrMnFeNi is -110.4 J/mol·K. This value and the values reported for other HEAs [59,60] are within the range reported for various conventional hydrogen storage materials which lie between -110 to -130 J/mol·K [2,8-11]. These results indicate that although HEAs and their corresponding high-entropy hydrides have high configurational entropy, the difference in their entropy is similar to conventional hydrogen storage materials. An entropy change in the range of -110 to -130 J/mol·K makes only a change of $T\Delta S$ = -6 kJ/mol at room temperature which is not high compared to the enthalpy in Eq. 1. Therefore, it is more reasonable to model the enthalpy using ML rather than the entropy similar to conventional hydrogen storage materials [8,27,28].

Regarding the third question, the ML GPR algorithm was used in the present study because GPR shows better accuracy than the multivariate regression, decision tree, and random forest algorithms [8]. The better performance of GPR for enthalpy prediction is due to the flexibility to represent data and inherently quantify uncertainty in the prediction [33]. GPR was implemented using four different kernels (exponential, rational quadratic, squared exponential and Matern), two different methods (5-fold cross-validation and self-validation) and with/without removal of outliers. The self-validation method results in a higher accuracy because all datum points are used as the training set and the test set is the same as the training set [28]. Removing outliers further increases the accuracy of the self-validation method because the data with possibly large experimental errors are excluded [37-39]. All kernels after removing outliers in the self-validation method exhibit high accuracy with $R^2$ > 0.9. However, the most accurate model for the enthalpy calculation is achieved by the exponential kernel and the lowest accuracy is obtained by the squared exponential and Matern kernels. The squared exponential kernel is infinitely differentiable which implies a very smooth function, which is not as good as other kernels for predicting the enthalpy [33,34]. The slightly lower accuracy of the Matern kernel can be attributed to its capability to predict local features because of the incorporation of a modified Bessel function of the second kind, but it is not always appropriate for predicting non-local data [34]. The rational quadratic and exponential kernels are both appropriate for enthalpy prediction [28], but this study suggests that the exponential kernel is slightly more accurate. The high accuracy achieved with the developed models suggests the high potential of ML GPR for the design of high-entropy hydrogen storage materials.

## 5. Conclusion

In this study, the hydride formation enthalpy of HEAs for solid-state hydrogen storage has been obtained via experiments using the van't Hoff analysis, via ML models using 420 datum points from literature and the GPR algorithm, and via first-principles calculations using DFT. All these methods produce consistent results with only small deviations. Experiments confirm that the performance of HEAs for room-temperature hydrogen storage is due to the enthalpy rather than entropy, suggesting that the prediction of the enthalpy is the most critical point in the design of HEAs. The ML GPR algorithms using different kernels show high accuracy in predicting the formation enthalpy, particularly using the self-validation method, the exponential kernel and after the removal of outliers. The findings of this study can be used to design new HEAs with hydride formation enthalpies of -25 kJ/mol to -39 kJ/mol for hydrogen storage at room temperature.

## Acknowledgements

The author S.D. thanks the MEXT, Japan for a scholarship. This work is supported in part by grants-in-aid for scientific research from the MEXT, Japan (JP19H05176 & JP21H00150), in part by the European Research Council (ERC) under the European Union's Horizon 2020 Research and Innovation Programme (grant agreement No 865855), in part by the State of Baden-Württemberg through bwHPC, and in part by the Deutsche Forschungsgemeinschaft (DFG, German Research Foundation) through project number 519607530 and grant number INST 40/575-1 FUGG (JUSTUS 2 cluster).

## Appendix

The dataset with 420 datum points for $AB_2$-type hydrogen storage materials is shown in Table A1. The data were gathered from [9-11,57-60,66-135].

Table A1. Experimental hydride formation enthalpies for hydrogen storage materials with different compositions.

| Composition | Enthalpy (kJ/mol) | Ref. | Outliers |
|---|---|---|---|
| $Ti_{1.1}Mn_1Cr_1$ | -22.9 | [9] | |
| $Ti_{0.99}Zr_{0.11}Mn_1Cr_1$ | -25.5 | [9] | |
| $Ti_{0.935}Zr_{0.165}Mn_1Cr_1$ | -26.6 | [9] | |
| $Ti_{0.88}Zr_{0.22}Mn_1Cr_1$ | -28.3 | [9] | |
| $Ti_{0.935}Zr_{0.165}Mn_1Cr_{0.95}Mo_{0.05}$ | -26.2 | [9] | |
| $Ti_{0.935}Zr_{0.165}Mn_1Cr_{0.9}Mo_{0.1}$ | -23.7 | [9] | |
| $Ti_{0.935}Zr_{0.165}Mn_1Cr_{0.85}Mo_{0.15}$ | -21.7 | [9] | |
| $Ti_{0.935}Zr_{0.165}Mn_1Cr_{0.98}W_{0.02}$ | -26.3 | [9] | |
| $Ti_{0.935}Zr_{0.165}Mn_1Cr_{0.95}W_{0.05}$ | -24.3 | [9] | * |
| $Ti_{0.935}Zr_{0.165}Mn_1Cr_{0.9}W_{0.1}$ | -22.6 | [9] | * |
| $Zr_1Fe_{1.9}$ | -21.8 | [10] | |
| $Zr_1Fe_2$ | -21.2 | [10] | |
| $Zr_1Fe_{2.5}$ | -18.3 | [10] | |
| $Ti_1Cr_{1.8}$ | -21.3 | [11] | |
| $Ti_1Cr_{1.9}$ | -26.5 | [11] | |
| $Ti_1Mn_{0.5}Cr_{1.5}$ | -21.4 | [11] | |
| $Ti_1Mn_{0.75}Cr_{1.25}$ | -19.9 | [11] | |
| $Ti_1Mn_1Cr_1$ | -19.6 | [11] | |
| $Ti_1Zr_1Mn_1V_1Fe_1$ | -46.5 | [57] | |
| $Ti_1Zr_1Mn_1Cr_{0.5}V_1Fe_1$ | -39.72 | [57] | |
| $Ti_1Zr_1Mn_1Cr_{0.75}V_1Fe_1$ | -36.21 | [57] | |
| $Ti_1Zr_1Mn_1Cr_1V_1Fe_1$ | -31.83 | [57] | |
| $Ti_1Zr_1Mn_1Cr_{1.25}V_1Fe_1$ | -31.56 | [57] | |
| $Ti_1Zr_1Mn_1Cr_{1.5}V_1Fe_1$ | -30.78 | [57] | |
| $Ti_1Zr_1Mn_1Cr_2V_1Fe_1$ | -25.95 | [57] | |
| $Ti_1Zr_1Mn_1Cr_1V_1$ | -51.21 | [57] | |
| $Ti_1Zr_1Mn_1Cr_1V_1Fe_{0.5}$ | -54.28 | [57] | * |
| $Ti_1Zr_1Mn_1Cr_1V_1Fe_{1.5}$ | -28.98 | [57] | |
| $Ti_1Zr_1Mn_1Cr_1V_1Fe_2$ | -21.88 | [57] | |
| $Ti_1Zr_1Cr_1V_1Fe_1$ | -42.44 | [57] | |
| $Ti_1Zr_1Mn_{0.5}Cr_1V_1Fe_1$ | -39.09 | [57] | |
| $Ti_1Zr_1Mn_{0.75}Cr_1V_1Fe_1$ | -36.12 | [57] | |
| $Ti_1Zr_1Mn_{1.25}Cr_1V_1Fe_1$ | -31.81 | [57] | |
| $Ti_1Zr_1Mn_{1.5}Cr_1V_1Fe_1$ | -27.01 | [57] | |
| $Ti_1Zr_1Mn_2Cr_1V_1Fe_1$ | -24.71 | [57] | |
| $Zr_1Mn_1Cr_1V_1Fe_1$ | -21.94 | [57] | |
| $Ti_{0.5}Zr_1Mn_1Cr_1V_1Fe_1$ | -25.63 | [57] | |
| $Ti_{1.5}Zr_1Mn_1Cr_1V_1Fe_1$ | -37.87 | [57] | |
| $Ti_2Zr_1Mn_1Cr_1V_1Fe_1$ | -44.03 | [57] | |
| $Ti_1Zr_1Mn_1Cr_1Fe_1$ | -42.46 | [57] | * |
| $Ti_1Zr_1Mn_1Cr_1V_{0.5}Fe_1$ | -41.02 | [57] | |
| $Ti_1Zr_1Mn_1Cr_1V_{1.5}Fe_1$ | -35.08 | [57] | |
| $Ti_1Zr_1Mn_1Cr_1V_2Fe_1$ | -36.59 | [57] | |
| $Ti_1Zr_{0.5}Mn_1Cr_1V_1Fe_1$ | -14.81 | [57] | |
| $Ti_1Zr_{1.5}Mn_1Cr_1V_1Fe_1$ | -46.17 | [57] | |
| $Ti_1Zr_2Mn_1Cr_1V_1Fe_1$ | -52.48 | [57] | |
| $Ti_{0.97}Zr_{0.019}Mn_{1.5}Cr_{0.045}V_{0.439}Al_{0.026}Fe_{0.097}$ | -25.4 | [59] | |
| $Ti_{0.98}Zr_{0.02}Mn_{1.5}Cr_{0.05}V_{0.43}Fe_{0.09}$ | -27.4 | [59] | |
| $Ti_{0.98}Zr_{0.02}Mn_{1.46}Cr_{0.05}V_{0.41}Fe_{0.08}$ | -29.2 | [59] | |
| $Ti_{0.95}Zr_{0.05}Mn_{1.48}V_{0.43}Al_{0.01}Fe_{0.08}$ | -29.4 | [59] | |
| $Ti_{0.955}Zr_{0.045}Mn_{1.52}V_{0.43}Al_{0.03}Fe_{0.12}$ | -28.7 | [59] | |
| $Ti_{12}Zr_{21.5}Mn_{8.1}Co_8Cr_{7.5}V_{10}Ni_{32.2}Sn_{0.3}Al_{0.4}$ | -33.7 | [60] | |
| $Ti_{12}Zr_{21.5}Mn_{8.1}Co_7Cr_{7.5}V_{10}Ni_{32.2}Sn_{0.3}Al_{0.4}Fe_1$ | -39.2 | [60] | |
| $Ti_{12}Zr_{21.5}Mn_{8.1}Co_6Cr_{7.5}V_{10}Ni_{32.2}Sn_{0.3}Al_{0.4}Fe_2$ | -38.4 | [60] | |
| $Ti_{12}Zr_{21.5}Mn_{8.1}Co_5Cr_{7.5}V_{10}Ni_{32.2}Sn_{0.3}Al_{0.4}Fe_3$ | -37.4 | [60] | |
| $Ti_{12}Zr_{21.5}Mn_{8.1}Co_4Cr_{7.5}V_{10}Ni_{32.2}Sn_{0.3}Al_{0.4}Fe_4$ | -38.5 | [60] | |
| $Ti_{12}Zr_{21.5}Mn_{8.1}Co_3Cr_{7.5}V_{10}Ni_{32.2}Sn_{0.3}Al_{0.4}Fe_5$ | -39.3 | [60] | |
| $Ti_1Mn_{0.85}Cr_1V_{0.05}Fe_{0.1}$ | -21.1 | [66] | |
| $Ti_1Mn_{0.7}Cr_1V_{0.1}Fe_{0.2}$ | -20.6 | [66] | |
| $Ti_1Mn_{0.55}Cr_1V_{0.15}Fe_{0.3}$ | -20.9 | [66] | |
| $Ti_1Mn_{0.4}Cr_1V_{0.2}Fe_{0.4}$ | -22 | [66] | |
| $Ti_{0.95}Zr_{0.05}Mn_1Cr_1$ | -20.8 | [66] | |
| $Ti_{0.95}Zr_{0.05}Mn_{0.8}Cr_{1.2}$ | -21.9 | [66] | |
| $Ti_{0.85}Zr_{0.15}Mn_1Cr_1$ | -25.3 | [66] | |
| $Ti_{0.85}Zr_{0.15}Mn_{0.8}Cr_{1.2}$ | -26.2 | [66] | |
| $Ti_9Zr_{28}Mn_{15}Co_5Cr_8Ni_{35}$ | -35.5 | [67] | |
| $Ti_{10}Zr_{27}Mn_{15}Co_5Cr_8Ni_{35}$ | -31.5 | [67] | |
| $Ti_{11}Zr_{26}Mn_{15}Co_5Cr_8Ni_{35}$ | -32.1 | [67] | |
| $Ti_{12}Zr_{25}Mn_{15}Co_5Cr_8Ni_{35}$ | -33.9 | [67] | |
| $Ti_{13}Zr_{24}Mn_{15}Co_5Cr_8Ni_{35}$ | -32.7 | [67] | |
| $Ti_{10}Zr_{28}Mn_{15}Co_5Cr_8Ni_{34}$ | -36.7 | [67] | |
| $Ti_{10}Zr_{26}Mn_{15}Co_5Cr_8Ni_{36}$ | -30.8 | [67] | |
| $Ti_{10}Zr_{25}Mn_{15}Co_5Cr_8Ni_{37}$ | -27.3 | [67] | |
| $Ti_{10}Zr_{24}Mn_{15}Co_5Cr_8Ni_{38}$ | -24.8 | [67] | |
| $Zr_{1.04}Al_{0.1}Fe_{1.9}$ | -15.2 | [68] | |
| $Zr_{1.04}Al_{0.15}Fe_{1.85}$ | -15.9 | [68] | |
| $Zr_{1.04}Al_{0.2}Fe_{1.8}$ | -18.1 | [68] | |
| $Zr_{1.04}Al_{0.25}Fe_{1.75}$ | -19.3 | [68] | * |
| $Zr_{1.04}Al_{0.3}Fe_{1.7}$ | -21.4 | [68] | * |
| $Zr_{1.04}V_{0.1}Fe_{1.9}$ | -23 | [68] | |
| $Zr_{1.04}V_{0.2}Fe_{1.8}$ | -24.2 | [68] | |
| $Zr_{1.04}V_{0.3}Fe_{1.7}$ | -29.1 | [68] | |
| $Zr_{1.04}V_{0.5}Fe_{1.5}$ | -42.5 | [68] | |
| $Ti_{0.104}Zr_{0.936}V_{0.3}Fe_{1.7}$ | -28.5 | [68] | |
| $Ti_{0.312}Zr_{0.728}V_{0.3}Fe_{1.7}$ | -26.1 | [68] | |
| $Ti_{0.52}Zr_{0.52}V_{0.3}Fe_{1.7}$ | -21.9 | [68] | |
| $Ti_{0.312}Zr_{0.728}V_{0.2}Fe_{1.8}$ | -23.5 | [68] | |
| $Ti_{0.312}Zr_{0.728}V_{0.5}Fe_{1.5}$ | -33.1 | [68] | |
| $Ti_{0.515}Zr_{0.485}Mn_{1.2}Cr_{0.8}$ | -33 | [69] | |
| $Ti_{0.515}Zr_{0.485}Mn_{1.2}Cr_{0.8}Fe_{0.1}$ | -30 | [69] | |
| $Ti_{0.515}Zr_{0.485}Mn_{1.2}Co_{0.1}Cr_{0.8}$ | -29 | [69] | |
| $Ti_{0.515}Zr_{0.485}Mn_{1.2}Cr_{0.8}Ni_{0.1}$ | -31 | [69] | |
| $Ti_{12}Zr_{21.5}Mn_{8.1}Co_8Cr_{7.5}V_{10}Ni_{32.2}Sn_{0.3}Al_{0.4}$ | -33.7 | [70] | |
| $Ti_{12}Zr_{21.5}Mn_{8.1}Co_7Cr_{7.5}V_{10}Ni_{32.2}Sn_{0.3}Al_{0.4}Mo_1$ | -35.4 | [70] | |
| $Ti_{12}Zr_{21.5}Mn_{8.1}Co_{6}Cr_{7.5}V_{10}Ni_{32.2}Sn_{0.3}Al_{0.4}Mo_{2}$ | -36.4 | [70] | |
| $Ti_{12}Zr_{21.5}Mn_{8.1}Co_{5}Cr_{7}.5V_{10}Ni_{32}.2Sn_{0}.3Al_{0.4}Mo_{3}$ | -36.7 | [70] | |
| $Ti_{12}Zr_{21.5}Mn_{8.1}Co_{4}Cr_{7.5}V_{10}Ni_{32.2}Sn_{0.3}Al_{0.4}Mo_{4}$ | -37.8 | [70] | |
| $Ti_{12}Zr_{21.5}Mn_{8.1}Co_{3}Cr_{7.5}V_{10}Ni_{32.2}Sn_{0.3}Al_{0.4}Mo_{5}$ | -38.7 | [70] | |
| $Ti_{1.02}Mn_{0.3}Cr_{1.1}Fe_{0.6}$ | -16.47 | [71] | |
| $Ti_{1.02}Mn_{0.3}Cr_{1.1}Fe_{0.6}La_{0.03}$ | -16.63 | [71] | |
| $Ti_{1.02}Mn_{0.3}Cr_{1.1}Fe_{0.6}Ce_{0.03}Y$ | -19.27 | [71] | |
| $Ti_{1.02}Mn_{0.3}Cr_{1.1}Fe_{0.6}Ho_{0.03}$ | -19.39 | [71] | * |
| $Ti_{5}Zr_{30}Mn_{19}Co_{5}Cr_{9}Ni_{32}$ | -33.9 | [72] | |
| $Ti_{5}Zr_{30}Mn_{19}Co_{5}Cr_{9}Ni_{31.5}B_{0.5}$ | -35.7 | [72] | |
| $Ti_{5}Zr_{30}Mn_{19}Co_{5}Cr_{9}Ni_{31}B_{1}$ | -34.7 | [72] | |
| $Ti_{5}Zr_{30}Mn_{19}Co_{5}Cr_{9}Ni_{30}B_{2}$ | -32.9 | [72] | |
| $Ti_{5}Zr_{30}Mn_{19}Co_{5}Cr_{9}Ni_{31.5}Fe_{0.5}$ | -36 | [72] | |
| $Ti_{5}Zr_{30}Mn_{19}Co_{5}Cr_{9}Ni_{31.5}Gd_{0.5}$ | -38.4 | [72] | * |
| $Ti_{5}Zr_{30}Mn_{19}Co_{5}Cr_{9}Ni_{30}Mg_{2}$ | -38.4 | [72] | |
| $Ti_{5}Zr_{30}Mn_{19}Co_{5}Cr_{9}Ni_{30}C_{0.7}$ | -31.3 | [72] | * |
| $Ti_{12}Zr_{22.8}Mn_{8.1}Co_{7}Cr_{7.5}V_{10}Ni_{32.2}Al_{0.4}$ | -41.6 | [73] | |
| $Ti_{12}Zr_{21.8}Mn_{8.1}Co_{7}Cr_{7.5}V_{10}Ni_{32.2}Al_{0.4}La_{1}$ | -41.4 | [73] | |
| $Ti_{12}Zr_{20.8}Mn_{8.1}Co_{7}Cr_{7}.5V_{10}Ni_{32.2}Al_{0.4}La_{2}$ | -40.6 | [73] | |
| $Ti_{12}Zr_{19.8}Mn_{8.1}Co_{7}Cr_{7.5}V_{10}Ni_{32.2}Al_{0.4}La_{3}$ | -40 | [73] | |
| $Ti_{12}Zr_{18.8}Mn_{8.1}Co_{7}Cr_{7.5}V_{10}Ni_{32.2}Al_{0.4}La_{4}$ | -36.6 | [73] | * |
| $Ti_{12}Zr_{17.8}Mn_{8.1}Co_{7}Cr_{7.5}V_{10}Ni_{32.2}Al_{0.4}La_{5}$ | -35.8 | [73] | * |
| $Ti_{0.7}Zr_{0.3}Mn_{1.8}V_{0.2}$ | -13 | [74] | |
| $Ti_{0.7}Zr_{0.3}Mn_{1.6}V_{0.4}$ | -14 | [74] | |
| $Ti_{0.7}Zr_{0.3}Mn_{1.4}V_{0.6}$ | -17 | [74] | |
| $Ti_{0.7}Zr_{0.3}Mn_{1.2}V_{0.8}$ | -20.5 | [74] | |
| $Ti_{0.7}Zr_{0.3}Mn_{1}V_{1}$ | -21 | [74] | |
| $Ti_{1}Co_{1.9}$ | -26.5 | [75] | * |
| $Ti_{1}Cr_{1.8}$ | -21.3 | [75] | |
| $Ti_{1}Cr_{1.75}Al_{0.05}$ | -24 | [75] | |
| $Ti_{1}Cr_{1.6}Al_{0.2}$ | -33.4 | [75] | |
| $Ti_{1}Cr_{1.7}Al_{0.2}$ | -28.1 | [75] | |
| $Ti_{1}Cr_{1.95}Al_{0.05}$ | -22.8 | [75] | |
| $Ti_{1}Mn_{0.2}Cr_{1.6}$ | -28 | [75] | |
| $Ti_{1}Mn_{0.95}Cr_{0.95}$ | -22.8 | [75] | |
| $Ti_{1}Cr_{1.8}B_{0.05}$ | -15.6 | [75] | * |
| $Ti_{1}Cr_{1.8}Si_{0.05}$ | -21.4 | [75] | |
| $Ti_{1}Cr_{1.75}Ni_{0.1}$ | -20.5 | [75] | |
| $Ti_{1}Cr_{1.7}Ni_{0.3}$ | -21.7 | [75] | |
| $Ti_{0.86}Cr_{1.9}Mo_{0.14}$ | -17.2 | [75] | |
| $Ti_{1}Cr_{1.9}$ | -24.8 | [75] | |
| $Ti_{1.1}Mn_{1}Cr_{1}$ | -22.9 | [76] | |
| $Ti_{1.045}Zr_{0.055}Mn_{1}Cr_{1}$ | -24.1 | [76] | |
| $Ti_{1.034}Zr_{0.066}Mn_1Cr_1$ | -22.1 | [76] | |
| $Ti_{0.99}Zr_{0.11}Mn_1Cr_1$ | -25.1 | [76] | |
| $Ti_1Mn_{0.2}Cr_{1.6}Fe_{0.2}$ | -19.66 | [77] | |
| $Ti_1Mn_{0.2}Cr_{1.58}Fe_{0.2}$ | -19.9 | [77] | |
| $Ti_1Mn_{0.2}Cr_{1.55}Fe_{0.2}$ | -20.32 | [77] | |
| $Ti_1Mn_{0.2}Cr_{1.5}Fe_{0.2}$ | -20.78 | [77] | |
| $Ti_1Mn_{0.25}Cr_{1.5}Fe_{0.25}$ | -19.32 | [77] | |
| $Ti_1Mn_{0.2}Cr_{1.5}Fe_{0.3}$ | -18.32 | [77] | |
| $Ti_{0.9}Zr_{0.1}Mn_{1.4}Cr_{0.35}V_{0.2}Fe_{0.05}$ | -25.89 | [77] | |
| $Ti_{0.45}Zr_{0.55}Mn_{0.09}V_{0.45}Ni_1$ | -41.7 | [78] | |
| $Ti_{0.45}Zr_{0.55}Mn_{0.19}V_{0.45}Ni_1$ | -39 | [78] | |
| $Ti_{0.45}Zr_{0.55}Mn_{0.28}V_{0.45}Ni_1$ | -36.9 | [78] | |
| $Ti_{0.45}Zr_{0.55}Mn_{0.23}V_{0.45}Ni_{0.85}$ | -47.6 | [78] | |
| $Ti_{0.45}Zr_{0.55}Mn_{0.33}V_{0.45}Ni_{0.85}$ | -44.2 | [78] | |
| $Ti_{0.45}Zr_{0.55}Mn_{0.53}V_{0.45}Ni_{0.85}$ | -36.4 | [78] | |
| $Ti_{0.45}Zr_{0.55}Mn_{0.7}V_{0.45}Ni_{0.85}$ | -32.8 | [78] | |
| $Ti_{0.45}Zr_{0.55}Mn_{0.53}V_{0.45}Ni_{0.7}$ | -54.8 | [78] | * |
| $Ti_{0.45}Zr_{0.55}Mn_{0.71}V_{0.45}Ni_{0.7}$ | -35.6 | [78] | |
| $Ti_{0.45}Zr_{0.55}Mn_{0.9}V_{0.45}Ni_{0.7}$ | -32.3 | [78] | |
| $Ti_{0.45}Zr_{0.55}Mn_{1.33}V_{0.45}Ni_{0.7}$ | -28.3 | [78] | |
| $Ti_{0.65}Zr_{0.35}Mn_1Cr_{0.8}Fe_{0.2}$ | -29 | [79] | |
| $Ti_{0.6825}Zr_{0.3675}Mn_1Cr_{0.8}Fe_{0.2}$ | -29 | [79] | |
| $Ti_{0.69875}Zr_{0.37625}Mn_1Cr_{0.8}Fe_{0.2}$ | -26 | [79] | |
| $Ti_{0.715}Zr_{0.385}Mn_1Cr_{0.8}Fe_{0.2}$ | -28 | [79] | |
| $Ti_{12}Zr_{21.5}Mn_{13.6}Cr_{4.5}V_{10}Ni_{38.1}Sn_{0.3}$ | -34 | [80] | |
| $Ti_{12}Zr_{21.5}Mn_{13.6}Co_{0.5}Cr_{4.5}V_{10}Ni_{37.6}Sn_{0.3}$ | -29.8 | [80] | |
| $Ti_{12}Zr_{21.5}Mn_{13.6}Co_1Cr_{4.5}V_{10}Ni_{37.1}Sn_{0.3}$ | -39.8 | [80] | |
| $Ti_{12}Zr_{21.5}Mn_{13.6}Co_{1.5}Cr_{4.5}V_{10}Ni_{36.6}Sn_{0.3}$ | -33.2 | [80] | |
| $Ti_{12}Zr_{21.5}Mn_{13.6}Co_2Cr_{4.5}V_{10}Ni_{36.1}Sn_{0.3}$ | -35.9 | [80] | |
| $Ti_{12}Zr_{21.5}Mn_{13.6}Co_{2.5}Cr_{4.5}V_{10}Ni_{35.6}Sn_{0.3}$ | -24.3 | [80] | |
| $Ti_{27}Zr_3Mn_{57}V_{13}$ | -30 | [81] | |
| $Ti_{32}Zr_4Mn_{59}V_5$ | -29 | [81] | |
| $Ti_{31}Zr_4Mn_{55}V_{10}$ | -31 | [81] | |
| $Ti_{28}Zr_3Mn_{49}V_{20}$ | -32 | [81] | |
| $Ti_{32}Zr_4Mn_{50}V_{14}$ | -33 | [81] | |
| $Ti_{36}Zr_4Mn_{55}V_5$ | -34 | [81] | |
| $Ti_{34}Zr_5Mn_{50}V_{11}$ | -35 | [81] | |
| $Ti_{32}Zr_4Mn_{46}V_{18}$ | -37 | [81] | |
| $Ti_{36}Zr_4Mn_{50}V_{10}$ | -38 | [81] | |
| $Ti_{40}Zr_6Mn_{49}V_5$ | -39 | [81] | |
| $Ti_{12}Zr_{21.5}Mn_{8.1}Cr_{7.5}V_{10}Ni_{32.2}Sn_{0.3}Al_{0.4}$ | -34.8 | [82] | |
| $Ti_{12}Zr_{21.5}Mn_{8.1}Cr_{7.5}V_{10}Ni_{32.2}Sn_{0.3}Al_{0.4}Cu_1$ | -29.8 | [82] | |
| $Ti_{12}Zr_{21.5}Mn_{8.1}Cr_{7.5}V_{10}Ni_{32.2}Sn_{0.3}Al_{0.4}Cu_2$ | -31.4 | [82] | |
| $Ti_{12}Zr_{21.5}Mn_{8.1}Cr_{7.5}V_{10}Ni_{32.2}Sn_{0.3}Al_{0.4}Cu_{3}$ | -31.7 | [82] | |
| $Ti_{12}Zr_{21.5}Mn_{8.1}Cr_{7.5}V_{10}Ni_{32.2}Sn_{0.3}Al_{0.4}Cu_{4}$ | -32.1 | [82] | |
| $Ti_{12}Zr_{21.5}Mn_{8.1}Cr_{7.5}V_{10}Ni_{32.2}Sn_{0.3}Al_{0.4}Cu_{5}$ | -31.5 | [82] | |
| $Ti_{13.3}Zr_{22.4}Mn_{5.4}Co_{1.5}Cr_{8.2}V_{9.7}Ni_{38.9}Sn_{0.3}Al_{0.4}$ | -29 | [83] | |
| $Ti_{12.9}Zr_{21.7}Mn_{5.5}Co_{1.5}Cr_{8.4}V_{9.8}Ni_{39.5}Sn_{0.3}Al_{0.4}$ | -28 | [83] | |
| $Ti_{12.5}Zr_{21}Mn_{5.6}Co_{1.5}Cr_{8.5}V_{10}Ni_{40.2}Sn_{0.3}Al_{0.4}$ | -24 | [83] | |
| $Ti_{12}Zr_{20.2}Mn_{5.7}Co_{1.5}Cr_{8.7}V_{10.2}Ni_{41}Sn_{0.3}Al_{0.4}$ | -23 | [83] | |
| $Ti_{11.7}Zr_{19.6}Mn_{5.8}Co_{1.6}Cr_{8.8}V_{10.3}Ni_{41.6}Sn_{0.3}Al_{0.4}$ | -21 | [83] | |
| $Ti_{13.3}Zr_{22.4}Mn_{4.9}Co_{4.8}Cr_{5.3}V_{9.7}Ni_{38.9}Sn_{0.3}Al_{0.4}$ | -26 | [83] | |
| $Ti_{12.9}Zr_{21.7}Mn_{5}Co_{4.9}Cr_{5.4}V_{9.8}Ni_{39.5}Sn_{0.3}Al_{0.4}$ | -24 | [83] | |
| $Ti_{12.5}Zr_{21}Mn_{5.1}Co_{5}Cr_{5.5}V_{10}Ni_{40.2}Sn_{0.3}Al_{0.4}$ | -22 | [83] | |
| $Ti_{12}Zr_{20.2}Mn_{5.2}Co_{5.1}Cr_{5.6}V_{10.2}Ni_{41}Sn_{0.3}Al_{0.4}$ | -20 | [83] | |
| $Ti_{11.7}Zr_{19.6}Mn_{5.3}Co_{5.2}Cr_{5.7}V_{10.3}Ni_{41.6}Sn_{0.3}Al_{0.4}$ | -14 | [83] | |
| $Ti_{13.3}Zr_{22.4}Mn_{4}Co_{7.7}Cr_{3.4}V_{9.7}Ni_{38.9}Sn_{0.3}Al_{0.4}$ | -28 | [83] | |
| $Ti_{12.9}Zr_{21.7}Mn_{4}Co_{7.9}Cr_{3.4}V_{9.8}Ni_{39.5}Sn_{0.3}Al_{0.4}$ | -27 | [83] | |
| $Ti_{12.5}Zr_{21}Mn_{4.1}Co_{8}Cr_{3.5}V_{10}Ni_{40.2}Sn_{0.3}Al_{0.4}$ | -12 | [83] | |
| $Ti_{12}Zr_{20.2}Mn_{4.2}Co_{8.2}Cr_{3.6}V_{10.2}Ni_{41}Sn_{0.3}Al_{0.4}$ | -11 | [83] | |
| $Ti_{11.7}Zr_{19.6}Mn_{4.2}Co_{8.3}Cr_{3.6}V_{10.3}Ni_{41.6}Sn_{0.3}Al_{0.4}$ | -7 | [83] | |
| $Ti_{31.35}Zr_{0.75}Mn_{48.6}Cr_{1.6}V_{13.95}Al_{0.05}Fe_{2.85}Si_{0.05}$ | -26.09 | [84] | |
| $Ti_{31.56}Zr_{0.7}Mn_{47.95}Cr_{1.5}V_{13.7}Al_{0.25}Fe_{2.85}Si_{0.15}$ | -27.09 | [84] | |
| $Ti_{31.3}Zr_{0.75}Mn_{48.4}Cr_{1.6}V_{14}Al_{0.01}Fe_{2.85}Si_{0.05}$ | -26.24 | [84] | |
| $Ti_{31.95}Zr_{0.7}Mn_{48.9}Cr_{1.5}V_{13.9}Al_{0.3}Fe_{1.55}Si_{0.15}$ | -27.05 | [84] | |
| $Ti_{0.52}Mn_{1}V_{0.39}Nb_{0.09}$ | -23.68 | [85] | * |
| $Ti_{0.51}Mn_{1}Cr_{0.1}V_{0.39}$ | -27.73 | [85] | |
| $Ti_{0.51}Mn_{1}V_{0.4}Mo_{0.09}$ | -19.81 | [85] | |
| $Ti_{0.52}Mn_{1}V_{0.43}Nb_{0.05}$ | -25.71 | [85] | |
| $Ti_{0.52}Mn_{1}V_{0.39}Nb_{0.09}$ | -23.68 | [85] | * |
| $Ti_{12.5}Zr_{21}Mn_{2.6}Co_{1.5}Cr_{8.5}V_{10}Ni_{43.9}$ | -24.9 | [86] | |
| $Ti_{12.5}Zr_{21}Mn_{5.6}Co_{1.5}Cr_{8.5}V_{10}Ni_{40.9}$ | -31.4 | [86] | |
| $Ti_{12.5}Zr_{21}Mn_{8.6}Co_{1.5}Cr_{8.5}V_{10}Ni_{37.9}$ | -33.8 | [86] | |
| $Ti_{12.5}Zr_{21}Mn_{11.6}Co_{1.5}Cr_{8.5}V_{10}Ni_{34.9}$ | -38.6 | [86] | |
| $Ti_{12.5}Zr_{21}Mn_{14.6}Co_{1.5}Cr_{8.5}V_{10}Ni_{31.9}$ | -39.3 | [86] | |
| $Ti_{12.5}Zr_{21}Mn_{17.6}Co_{1.5}Cr_{8.5}V_{10}Ni_{28.9}$ | -40 | [86] | |
| $Ti_{12.5}Zr_{21}Mn_{20.6}Co_{1.5}Cr_{8.5}V_{10}Ni_{25.9}$ | -40.3 | [86] | |
| $Ti_{12.5}Zr_{21}Mn_{23.6}Co_{1.5}Cr_{8.5}V_{10}Ni_{22.9}$ | -42.3 | [86] | |
| $Ti_{12}Zr_{21.5}Mn_{8.1}Co_{8}Cr_{7.5}V_{10}Ni_{32.2}Sn_{0.3}Al_{0.4}$ | -33.7 | [87] | |
| $Ti_{12}Zr_{21.5}Mn_{8.1}Co_{8}Cr_{7.5}V_{10}Ni_{31.2}Sn_{0.3}Al_{0.4}Si_{1}$ | -38.3 | [87] | |
| $Ti_{12}Zr_{21.5}Mn_{8.1}Co_{8}Cr_{7.5}V_{10}Ni_{30.2}Sn_{0.3}Al_{0.4}Si_{2}$ | -40.3 | [87] | |
| $Ti_{12}Zr_{21.5}Mn_{8.1}Co_{8}Cr_{7.5}V_{10}Ni_{29.2}Sn_{0.3}Al_{0.4}Si_{3}$ | -40.7 | [87] | * |
| $Ti_{12}Zr_{21.5}Mn_{8.1}Co_{8}Cr_{7.5}V_{10}Ni_{28.2}Sn_{0.3}Al_{0.4}Si_{4}$ | -42.6 | [87] | * |
| $Ti_{12}Zr_{21.5}Mn_{8.1}Co_{8}Cr_{7.5}V_{10}Ni_{32.2}Sn_{0.3}Al_{0.4}$ | -32 | [88] | |
| $Ti_{11.4}Zr_{21.5}Mn_{8.1}Co_{8}Cr_{7.5}V_{10}Ni_{31.8}Sn_{0.3}Al_{0.4}Mg_{1}$ | -32.2 | [88] | |
| $Ti_{10.8}Zr_{21.5}Mn_{8.1}Co_{8}Cr_{7.5}V_{10}Ni_{31.4}Sn_{0.3}Al_{0.4}Mg_{2}$ | -36.2 | [88] | |
| $Ti_{10.2}Zr_{21.5}Mn_{8.1}Co_{8}Cr_{7.5}V_{10}Ni_{31}Sn_{0.3}Al_{0.4}Mg_{3}$ | -35.2 | [88] | * |
| $Ti_{9.6}Zr_{21.5}Mn_{8.1}Co_{8}Cr_{7.5}V_{10}Ni_{30.6}Sn_{0.3}Al_{0.4}Mg_{4}$ | -31.2 | [88] | * |
| $Ti_{12}Zr_{21.5}Mn_{8.1}Co_{8}Cr_{7.5}V_{10}Ni_{32.2}Sn_{0.3}Al_{0.4}$ | -31 | [89] | |
| $Ti_{12.1}Zr_{21.6}Mn_{8.2}Co_{7}Cr_{7.6}V_{10.1}Ni_{32.4}Sn_{0.3}Al_{0.4}Y_{0.3}$ | -32 | [89] | |
| $Ti_{12.2}Zr_{21.8}Mn_{8.2}Co_{6.1}Cr_{7.6}V_{10.1}Ni_{32.6}Sn_{0.3}Al_{0.4}Y_{0.6}$ | -34 | [89] | |
| $Ti_{12.3}Zr_{22}Mn_{8.3}Co_{5.1}Cr_{7.7}V_{10.2}Ni_{32.9}Sn_{0.3}Al_{0.4}Y_{1}$ | -32 | [89] | |
| $Ti_{12.3}Zr_{22.1}Mn_{8.3}Co_{4.1}Cr_{7.7}V_{10.3}Ni_{33.1}Sn_{0.3}Al_{0.4}Y_{1.3}$ | -31 | [89] | |
| $Ti_{12}Zr_{21.5}Mn_{5.6}Co_{1.5}Cr_{8.5}V_{10}Ni_{40.2}Sn_{0.3}Al_{0.4}$ | -34.41 | [90] | |
| $Ti_{12}Zr_{21.5}Mn_{5.1}Co_{5}Cr_{5.5}V_{10}Ni_{40.2}Sn_{0.3}Al_{0.4}$ | -32.17 | [90] | |
| $Ti_{12}Zr_{21.5}Mn_{4.1}Co_{8}Cr_{3.5}V_{10}Ni_{40.2}Sn_{0.3}Al_{0.4}$ | -31.15 | [90] | |
| $Ti_{4.5}Zr_{36}Mn_{1.3}Co_{0.4}Cr_{0.7}V_{0.8}Ni_{39.5}Sn_{16.5}Al_{0.4}$ | -33.43 | [90] | * |
| $Ti_{9.1}Zr_{35.8}Mn_{2.2}Co_{2.5}Cr_{1.5}V_{3.7}Ni_{35.7}Sn_{9.2}Al_{0.4}$ | -33.41 | [90] | * |
| $Ti_{8.5}Zr_{30.5}Mn_{1.9}Co_{3.9}Cr_{0.7}V_{3.8}Ni_{38.8}Sn_{11.5}Al_{0.4}$ | -32.69 | [90] | * |
| $Ti_{12}Zr_{21.5}Mn_{13.5}Cr_{4.5}V_{10}Ni_{37.7}Sn_{0.3}Al_{0.5}$ | -36.38 | [91] | |
| $Ti_{12}Zr_{21.5}Mn_{4.1}Co_{8}Cr_{3.5}V_{10}Ni_{40.2}Sn_{0.3}Al_{0.4}$ | -30.59 | [91] | |
| $Ti_{14.6}Zr_{19.4}Mn_{7.5}Co_{6.4}Cr_{5.1}V_{8.6}Ni_{38.3}Sn_{0.1}$ | -32 | [92] | |
| $Ti_{6.5}Zr_{24.6}Mn_{22.2}V_{3.7}Ni_{38.6}Fe_{4.4}$ | -31.8 | [92] | |
| $Ti_{6.5}Zr_{24.8}Mn_{22.6}V_{3.8}Ni_{37.9}Fe_{4.2}La_{0.1}$ | -35.4 | [92] | |
| $Ti_{0.24}Zr_{0.76}Mn_{0.64}V_{0.14}Ni_{1.2}Fe_{0.18}$ | -28.43 | [93] | |
| $Ti_{0.24}Zr_{0.76}Mn_{0.64}V_{0.14}Ni_{1.2}Fe_{0.18}La_{0.01}$ | -29.42 | [93] | |
| $Ti_{0.1}Zr_{0.9}Mn_{0.9}V_{0.1}Ni_{0.5}Fe_{0.5}$ | -30.58 | [94] | |
| $Ti_{0.5}Zr_{0.5}V_{0.4}Ni_{0.4}Fe_{1.2}$ | -28.33 | [95] | |
| $Ti_{0.25}Zr_{0.75}V_{0.2}Ni_{0.8}Fe_{1}$ | -23.96 | [95] | |
| $Ti_{0.9}Zr_{0.1}Mn_{1.34}V_{0.3}$ | -28.7 | [96] | |
| $Ti_{0.9}Zr_{0.1}Mn_{1.47}Cr_{0.4}V_{0.2}$ | -26.8 | [96] | |
| $Ti_{12}Zr_{21.5}Mn_{8.1}Co_{8}Cr_{7.5}V_{10}Ni_{32.2}Sn_{0.3}Al_{0.4}$ | -32.8 | [97] | |
| $Ti_{12}Zr_{21.5}Mn_{8.1}Co_{8}Cr_{7.5}V_{9.6}Ni_{31.6}Sn_{0.3}Al_{0.4}B_{1}$ | -38.9 | [97] | |
| $Ti_{12}Zr_{21.5}Mn_{8.1}Co_{8}Cr_{7.5}V_{9.2}Ni_{31}Sn_{0.3}Al_{0.4}B_{2}$ | -37.5 | [97] | |
| $Ti_{12}Zr_{21.5}Mn_{8.1}Co_{8}Cr_{7.5}V_{8.8}Ni_{30.4}Sn_{0.3}Al_{0.4}B_{3}$ | -35.3 | [97] | |
| $Ti_{12}Zr_{21.5}Mn_{8.1}Co_{8}Cr_{7.5}V_{8.4}Ni_{29.8}Sn_{0.3}Al_{0.4}B_{4}$ | -35.5 | [97] | * |
| $Ti_{12}Zr_{21.5}Mn_{8.1}Co_{8}Cr_{7.5}V_{8}Ni_{29.2}Sn_{0.3}Al_{0.4}B_{5}$ | -42.9 | [97] | * |
| $Ti_{12}Zr_{21.5}Mn_{13.6}Co_{1.5}Cr_{8.5}V_{10}Ni_{32.2}Sn_{0.3}Al_{0.4}$ | -36.69 | [98] | |
| $Ti_{12}Zr_{21.5}Mn_{13.6}Co_{1.5}Cr_{4.5}V_{10}Ni_{36.2}Sn_{0.3}Al_{0.4}$ | -35.67 | [98] | |
| $Ti_{14}Zr_{19.5}Mn_{13.6}Co_{1.5}Cr_{4.5}V_{10}Ni_{36.2}Sn_{0.3}Al_{0.4}$ | -34.38 | [98] | |
| $Ti_{12}Zr_{21.5}Mn_{15.6}Co_{1.5}Cr_{6.5}V_{10}Ni_{32.2}Sn_{0.3}Al_{0.4}$ | -36.96 | [98] | |
| $Ti_{12}Zr_{21.5}Mn_{5.6}Co_{1.5}Cr_{8.5}V_{10}Ni_{40.2}Sn_{0.3}Al_{0.4}$ | -34.4 | [99] | |
| $Ti_{12}Zr_{20.5}Mn_{5.6}Co_{1.5}Cr_{8.5}V_{10}Ni_{40.2}Sn_{0.3}Al_{0.4}Y_{1}$ | -34.91 | [99] | |
| $Ti_{12}Zr_{19.5}Mn_{5.6}Co_{1.5}Cr_{8.5}V_{10}Ni_{40.2}Sn_{0.3}Al_{0.4}Y_{2}$ | -35.05 | [99] | |
| $Ti_{12}Zr_{18.5}Mn_{5.6}Co_{1.5}Cr_{8.5}V_{10}Ni_{40.2}Sn_{0.3}Al_{0.4}Y_{3}$ | -34.95 | [99] | |
| $Ti_{12}Zr_{17.5}Mn_{5.6}Co_{1.5}Cr_{8.5}V_{10}Ni_{40.2}Sn_{0.3}Al_{0.4}Y_{4}$ | -35.05 | [99] | * |
| $Ti_{12}Zr_{21.5}Mn_{4.1}Co_{8}Cr_{3.5}V_{10}Ni_{40.2}Sn_{0.3}Al_{0.4}$ | -31.15 | [99] | |
| $Ti_{12}Zr_{20.5}Mn_{4.1}Co_{8}Cr_{3.5}V_{10}Ni_{40.2}Sn_{0.3}Al_{0.4}Y_{1}$ | -32.17 | [99] | |
| $Ti_{12}Zr_{19.5}Mn_{4.1}Co_{8}Cr_{3.5}V_{10}Ni_{40.2}Sn_{0.3}Al_{0.4}Y_{2}$ | -32.9 | [99] | |
| $Ti_{12}Zr_{18.5}Mn_{4.1}Co_8Cr_{3.5}V_{10}Ni_{40.2}Sn_{0.3}Al_{0.4}Y_3$ | -32.17 | [99] | |
| $Ti_{12}Zr_{17.5}Mn_{4.1}Co_8Cr_{3.5}V_{10}Ni_{40.2}Sn_{0.3}Al_{0.4}Y_4$ | -31.87 | [99] | * |
| $Ti_{0.15}Zr_{0.85}Mn_{0.7}V_{0.12}Ni_{1.2}Fe_{0.12}La_{0.03}$ | -34 | [100] | |
| $Ti_{12.1}Zr_{21.3}Mn_{8.1}Co_8Cr_{7.5}V_{10.5}Ni_{31.7}Al_{0.8}$ | -36 | [101] | |
| $Ti_{8.1}Zr_{14.2}Mn_{5.4}Co_{5.3}Cr_5V_7Ni_{21.1}Al_{0.5}Li_{33.4}$ | -34.8 | [101] | * |
| $Ti_1Cr_{1.8}$ | -20.2 | [102] | |
| $Ti_{0.98}Zr_{0.02}Mn_{1.5}Cr_{0.05}V_{0.43}Fe_{0.09}$ | -27.4 | [102] | |
| $Ti_1Mn_{1.5}$ | -28.7 | [102] | |
| $Zr_1Cr_{0.5}Fe_{1.5}$ | -25.6 | [102] | |
| $Ti_1Mn_{1.4}V_{0.62}$ | -28.6 | [102] | |
| $Zr_1Mn_2$ | -53.2 | [102] | * |
| $Ti_{0.3}Zr_1Cr_{0.3}Ni_{1.4}$ | -30.39 | [103] | |
| $Ti_{0.4}Zr_{0.8}Cr_{0.4}Ni_{1.4}$ | -27.3 | [103] | |
| $Ti_{0.5}Zr_1Cr_{0.5}Ni_{1.2}$ | -30.22 | [103] | |
| $Ti_{0.3}Zr_{0.8}Mn_{0.1}Cr_{0.5}Ni_{1.1}$ | -30.81 | [103] | |
| $Ti_{0.15}Zr_{0.8}Cr_{0.15}V_{0.8}Ni_1$ | -31.16 | [103] | |
| $Ti_{0.5}Zr_{0.7}Cr_{0.5}Ni_{1.3}$ | -27.25 | [103] | |
| $Ti_{0.3}Zr_{0.5}Cr_{0.3}V_{1.2}Ni_{0.7}Cu_{0.1}$ | -30.43 | [103] | |
| $Ti_{0.5}Zr_{0.5}V_{0.7}Ni_{1.1}Cu_{0.2}$ | -26.67 | [103] | * |
| $Ti_{0.3}Zr_{0.5}Cr_{0.35}V_{0.8}Ni_1$ | -28.29 | [103] | |
| $Ti_{0.6}Zr_{0.5}V_{0.8}Ni_{1.1}$ | -27.85 | [103] | |
| $Ti_{0.7}Zr_{0.4}V_{0.6}Ni_{1.3}$ | -25.2 | [103] | |
| $Zr_1Mn_2$ | -39.69 | [104] | |
| $Zr_1Fe_2$ | -29.14 | [105] | |
| $Ti_{0.2}Zr_{0.8}Cr_{0.2}V_{0.3}Fe_{1.5}$ | -31.85 | [105] | |
| $Zr_1Mn_1Fe_1$ | -30.94 | [106] | |
| $Zr_1Mn_{1.8}Fe_{0.2}$ | -32.74 | [106] | |
| $Ti_{0.8}Zr_{0.2}Mn_{1.8}Mo_{0.2}$ | -29 | [107] | * |
| $Ti_{0.9}Zr_{0.1}Mn_{1.4}Cr_{0.4}V_{0.2}$ | -31.42 | [107] | |
| $Ti_{0.8}Zr_{0.2}Mn_{1.2}Cr_{0.4}$ | -33.55 | [107] | |
| $Zr_1V_{0.5}Fe_{1.5}$ | -31.81 | [108] | |
| $Zr_1Co_1Cr_1$ | -38.46 | [108] | |
| $Ti_{0.3}Zr_{0.7}Mn_2$ | -33 | [109] | |
| $Ti_{0.4}Zr_{0.6}Mn_2$ | -35.66 | [109] | |
| $Zr_1Cr_{0.4}Fe_{1.6}$ | -31.18 | [110] | |
| $Zr_1Cr_{0.5}Fe_{1.5}$ | -34.41 | [110] | |
| $Zr_1Cr_{0.6}Fe_{1.4}$ | -35.46 | [110] | |
| $Zr_1Cr_{0.8}Fe_{1.2}$ | -37.51 | [110] | |
| $Zr_1Cr_{0.9}Fe_{1.1}$ | -38.64 | [110] | |
| $Zr_1Cr_1Fe_1$ | -39.06 | [110] | |
| $Ti_{0.15}Zr_{0.85}Mn_{0.64}V_{0.11}Ni_{1.097}Fe_{0.11}La_{0.03}$ | -33.98 | [111] | |
| $Ti_{0.15}Zr_{0.85}Mn_{0.657}V_{0.113}Ni_{1.126}Fe_{0.113}La_{0.03}$ | -35.25 | [111] | |
| $Ti_{0.15}Zr_{0.85}Mn_{0.674}V_{0.116}Ni_{1.155}Fe_{0.116}La_{0.03}$ | -34.54 | [111] | |
| $Ti_{0.85}Zr_{0.15}Mn_{1.22}Cr_{0.2}V_{0.3}Ni_{0.22}Fe_{0.06}$ | -26.4 | [112] | |
| $Ti_{0.1}Zr_{0.9}Mn_{0.9}Co_{0.5}V_{0.1}Fe_{0.5}$ | -29.99 | [113] | |
| $Ti_{0.11}Zr_{0.99}Mn_{0.9}Co_{0.5}V_{0.1}Fe_{0.5}$ | -32.51 | [113] | |
| $Ti_{0.1}Zr_{0.9}Mn_{0.99}Co_{0.5}V_{0.11}Fe_{0.5}$ | -30.2 | [113] | |
| $Ti_{0.1}Zr_{0.9}Mn_{0.9}Co_{0.55}V_{0.1}Fe_{0.55}$ | -29.4 | [113] | |
| $Ti_{0.8}Zr_{0.2}Mn_{1.5}V_{0.4}Fe_{0.1}$ | -15.5 | [114] | |
| $Ti_{12}Zr_{21.5}Mn_{8.1}Co_{8}Cr_{7.5}V_{10}Ni_{32.2}Sn_{0.3}Al_{0.4}$ | -32 | [92] | |
| $Ti_{6.5}Zr_{25}Mn_{22.2}V_{3.9}Ni_{38}Fe_{3.8}La_{0.3}$ | -31.82 | [92] | |
| $Ti_{12}Zr_{22.8}Mn_{8.1}Co_{7}Cr_{7.5}V_{10}Ni_{32.2}Al_{0.4}$ | -41.6 | [115] | |
| $Ti_{12}Zr_{21.8}Mn_{8.1}Co_{7}Cr_{7.5}V_{10}Ni_{32.2}Al_{0.4}Ce_{1}$ | -40.2 | [115] | |
| $Ti_{12}Zr_{20.8}Mn_{8.1}Co_{7}Cr_{7.5}V_{10}Ni_{32.2}Al_{0.4}Ce_{2}$ | -40 | [115] | |
| $Ti_{12}Zr_{19.8}Mn_{8.1}Co_{7}Cr_{7.5}V_{10}Ni_{32.2}Al_{0.4}Ce_{3}$ | -39.1 | [115] | |
| $Ti_{12}Zr_{18.8}Mn_{8.1}Co_{7}Cr_{7.5}V_{10}Ni_{32.2}Al_{0.4}Ce_{4}$ | -38.3 | [115] | * |
| $Ti_{12}Zr_{17.8}Mn_{8.1}Co_{7}Cr_{7.5}V_{10}Ni_{32.2}Al_{0.4}Ce_{5}$ | -38.5 | [115] | * |
| $Ti_{0.95}Zr_{0.05}Mn_{0.7}Cr_{0.2}V_{0.4}Ni_{0.3}Fe_{0.5}$ | -25 | [116] | |
| $Ti_{0.9}Zr_{0.1}Mn_{0.4}Cr_{0.65}V_{0.2}Ni_{0.3}Fe_{0.46}$ | -25.29 | [116] | |
| $Ti_{0.85}Zr_{0.15}Mn_{0.8}Cr_{0.15}V_{0.35}Ni_{0.3}Fe_{0.42}$ | -21.17 | [116] | |
| $Ti_{0.85}Zr_{0.15}Mn_{1.1}Cr_{0.2}V_{0.3}Ni_{0.2}Fe_{0.21}$ | -26.61 | [116] | |
| $Ti_{0.65}Zr_{0.35}Mn_{0.81}Cr_{0.84}Ni_{0.3}Fe_{0.05}$ | -21.74 | [116] | |
| $Ti_{0.55}Zr_{0.45}Mn_{0.81}Cr_{0.84}Ni_{0.3}Fe_{0.84}$ | -18.46 | [116] | |
| $Ti_{0.8}Zr_{0.2}Mn_{0.7}Cr_{1.2}V_{0.1}$ | -26.8 | [117] | |
| $Ti_{0.8}Zr_{0.2}Mn_{0.75}Cr_{1.15}V_{0.1}$ | -27.12 | [117] | |
| $Ti_{0.8}Zr_{0.2}Mn_{0.8}Cr_{1.1}V_{0.1}$ | -27.8 | [117] | |
| $Ti_{0.8}Zr_{0.2}Mn_{1}Cr_{0.9}V_{0.1}$ | -27.6 | [117] | |
| $Ti_{0.96}Zr_{0.04}Mn_{0.8}Cr_{1.2}$ | -24.6 | [117] | |
| $Ti_{0.93}Zr_{0.07}Mn_{0.8}Cr_{1.2}$ | -23.9 | [117] | |
| $Ti_{0.9}Zr_{0.1}Mn_{0.8}Cr_{1.2}$ | -24.4 | [117] | |
| $Ti_{0.8}Zr_{0.2}Mn_{0.8}Cr_{1.2}$ | -25.6 | [117] | |
| $Ti_{0.55}Zr_{0.45}Mn_{0.386}Cr_{0.84}Ni_{0.2}Fe_{0.55}Cu_{0.028}$ | -28.3 | [118] | |
| $Ti_{0.85}Zr_{0.15}Mn_{1.33}V_{0.3}$ | -33 | [119] | |
| $Ti_{0.8}Zr_{0.2}Mn_{1.2}Cr_{0.6}V_{0.2}$ | -29 | [119] | |
| $Ti_{0.9}Zr_{0.1}Mn_{1.4}Cr_{0.4}V_{0.2}$ | -27 | [119] | |
| $Ti_{1.125}Zr_{0.125}Mn_{1.1}Cr_{0.85}Mo_{0.05}$ | -24 | [120] | |
| $Zr_{1}V_{0.2}Fe_{1.8}$ | -23.6 | [121] | |
| $Zr_{1}Cr_{0.2}Fe_{1.8}$ | -22.3 | [121] | |
| $Zr_{1}Mn_{0.2}Fe_{1.8}$ | -21.8 | [121] | |
| $Zr_{1}Co_{0.2}Fe_{1.8}$ | -16.8 | [121] | |
| $Zr_{1}Ni_{0.2}Fe_{1.8}$ | -21.5 | [121] | |
| $Zr_{1}Fe_{1.8}Cu_{0.2}$ | -19.6 | [121] | * |
| $Zr_{1}Fe_{1.8}Mo_{0.2}$ | -25.9 | [121] | * |
| $Ti_{0.7}Zr_{0.3}V_{0.6}Fe_{1.4}$ | -31.1 | [121] | |
| $Ti_{0.8}Zr_{0.2}V_{0.4}Fe_{1}$ | -19.7 | [121] | |
| $Ti_{0.9}Zr_{0.1}V_{0.3}Fe_{1.7}$ | -13.5 | [121] | |
| $Ti_{0.4}Zr_{0.6}Ni_{0.8}Fe_{1.2}$ | -16.5 | [121] | |
| $Ti_{0.2}Zr_{0.8}V_{0.2}Ni_{0.8}Fe_{1}$ | -26.8 | [121] | |
| $Ti_{0.9}Zr_{0.1}V_{0.2}Ni_{0.3}Fe_{1.5}$ | -12.1 | [121] | |
| $Ti_{0.8}Zr_{0.2}Mn_{1.2}Cr_{0.8}$ | -28.87 | [122] | |
| $Ti_{1}Cr_{1.9}$ | -22.58 | [123] | |
| $Ti_{1.02}Mn_{0.3}Cr_{1.1}Fe_{0.6}$ | -20.2 | [124] | |
| $Ti_{1.02}Mn_{0.3}Cr_{1.05}Fe_{0.65}$ | -19.8 | [124] | |
| $Ti_{1.02}Mn_{0.3}Cr_{1.03}Fe_{0.67}$ | -19.3 | [124] | |
| $Ti_{1.02}Mn_{0.3}Cr_{1}Fe_{0.7}$ | -19.9 | [124] | |
| $Ti_{1.02}Mn_{0.3}Cr_{0.95}Fe_{0.75}$ | -18.7 | [124] | |
| $Ti_{1.02}Mn_{0.25}Cr_{1}Fe_{0.75}$ | -19 | [124] | |
| $Ti_{1}Mn_{0.2}Cr_{1.2}Fe_{0.6}$ | -13.7 | [125] | |
| $Ti_{1.02}Mn_{0.2}Cr_{1.2}Fe_{0.6}$ | -16.67 | [125] | |
| $Ti_{1.05}Mn_{0.2}Cr_{1.2}Fe_{0.6}$ | -17.46 | [125] | |
| $Ti_{1.1}Mn_{0.2}Cr_{1.2}Fe_{0.6}$ | -14.98 | [125] | |
| $Zr_{1}Cr_{0.8}Fe_{1.2}$ | -28.91 | [126] | |
| $Zr_{1}Cr_{0.5}Fe_{1.5}$ | -24.23 | [126] | |
| $Ti_{0.05}Zr_{0.95}Cr_{0.8}Fe_{1.2}$ | -28.28 | [126] | |
| $Ti_{0.1}Zr_{0.9}Cr_{1}Fe_{1}$ | -31.38 | [126] | |
| $Ti_{0.1}Zr_{0.9}Cr_{0.9}Fe_{1.1}$ | -29.66 | [126] | |
| $Ti_{0.1}Zr_{0.9}Cr_{0.8}Fe_{1.2}$ | -28.24 | [126] | |
| $Ti_{0.1}Zr_{0.9}Cr_{0.7}Fe_{1.3}$ | -25.06 | [126] | |
| $Ti_{0.1}Zr_{0.9}Cr_{0.6}Fe_{1.4}$ | -24.56 | [126] | |
| $Ti_{0.15}Zr_{0.85}Cr_{0.8}Fe_{1.2}$ | -27.66 | [126] | |
| $Ti_{0.2}Zr_{0.8}Cr_{1}Fe_{1}$ | -29.46 | [126] | |
| $Ti_{0.2}Zr_{0.8}Cr_{0.8}Fe_{1.2}$ | -26.31 | [126] | |
| $Ti_{0.3}Zr_{0.7}Cr_{1}Fe_{1}$ | -32.01 | [126] | |
| $Ti_{0.9}Zr_{0.1}Mn_{1}Cr_{0.9}V_{0.1}$ | -25.54 | [127] | |
| $Ti_{0.95}Zr_{0.1}Mn_{1}Cr_{0.9}V_{0.1}$ | -25.91 | [127] | |
| $Ti_{1}Zr_{0.1}Mn_{1}Cr_{0.9}V_{0.1}$ | -26.51 | [127] | |
| $Ti_{1.05}Zr_{0.1}Mn_{1}Cr_{0.9}V_{0.1}$ | -27.79 | [127] | |
| $Ti_{0.85}Zr_{0.15}Mn_{1}Cr_{0.9}V_{0.1}$ | -27.53 | [127] | |
| $Ti_{0.8}Zr_{0.15}Mn_{1}Cr_{0.9}V_{0.1}$ | -26.05 | [127] | |
| $Ti_{0.75}Zr_{0.2}Mn_{1}Cr_{0.9}V_{0.1}Sn$ | -28.09 | [127] | |
| $Ti_{0.9}Zr_{0.1}Mn_{0.8}Cr_{1.2}$ | -24.76 | [128] | |
| $Ti_{0.85}Zr_{0.15}Mn_{0.8}Cr_{1.2}$ | -25.78 | [128] | |
| $Ti_{0.8}Zr_{0.2}Mn_{0.8}Cr_{1.2}$ | -27.41 | [128] | |
| $Ti_{0.75}Zr_{0.25}Mn_{0.8}Cr_{1.2}$ | -28.43 | [128] | |
| $Ti_{0.7875}Zr_{0.2625}Mn_{0.8}Cr_{1.2}$ | -29.54 | [128] | |
| $Ti_{0.825}Zr_{0.275}Mn_{0.8}Cr_{1.2}$ | -31.46 | [128] | |
| $Ti_{0.8625}Zr_{0.2875}Mn_{0.8}Cr_{1.2}$ | -32.31 | [128] | |
| $Ti_{0.21}Zr_{0.79}Mn_{1}Fe_{1}$ | -33 | [129] | |
| $Ti_{0.25}Zr_{0.75}Mn_{1.1}Fe_{0.9}$ | -29.4 | [130] | |
| $Ti_{0.2}Zr_{0.8}Co_1Fe_1$ | -31 | [131] | |
| $Ti_{0.2}Zr_{0.8}Mn_{0.25}Co_{0.75}Fe_1$ | -32.76 | [131] | |
| $Ti_{0.2}Zr_{0.8}Mn_{0.5}Co_{0.5}Fe_1$ | -30.71 | [131] | |
| $Ti_{0.2}Zr_{0.8}Mn_{0.75}Co_{0.25}Fe_1$ | -33.39 | [131] | |
| $Ti_{0.2}Zr_{0.8}Mn_{0.85}Co_{0.15}Fe_1$ | -31.51 | [131] | |
| $Ti_{0.2}Zr_{0.8}Mn_1$ | -29.62 | [131] | * |
| $Zr_1Cr_{0.6}SnFe_{1.4}$ | -27.29 | [132] | |
| $Ti_{0.2}Zr_{0.8}Cr_{0.6}Fe_{1.4}$ | -26.97 | [132] | |
| $Ti_{0.3}Zr_{0.7}Cr_{0.6}Fe_{1.4}$ | -22.53 | [132] | |
| $Zr_1Mn_1Cr_{0.25}Fe_1$ | -25.4 | [133] | |
| $Zr_1Cr_1Fe_{1.4}$ | -19.5 | [133] | |
| $Zr_1Cr_1Fe_{1.5}$ | -23.3 | [133] | |
| $Zr_{1.05}Mn_{0.2}Fe_{1.8}$ | -18.06 | [134] | |
| $Zr_{1.05}Mn_{0.3}Fe_{1.7}$ | -19.06 | [134] | |
| $Zr_{1.05}Mn_{0.4}Fe_{1.6}$ | -19.97 | [134] | |
| $Zr_{1.05}Mn_{0.6}Fe_{1.4}$ | -25.5 | [134] | |
| $Zr_{1.05}Mn_{0.2}V_{0.1}Fe_{1.7}$ | -19.27 | [134] | |
| $Zr_{1.05}Mn_{0.3}V_{0.1}Fe_{1.6}$ | -20.75 | [134] | |
| $Ti_{0.42}Zr_{0.63}Mn_{0.7}Fe_{1.3}$ | -24.57 | [134] | |
| $Ti_{0.42}Zr_{0.63}Mn_{0.8}Fe_{1.2}$ | -25.69 | [134] | |
| $Ti_{0.42}Zr_{0.63}Mn_{0.9}Fe_{1.1}$ | -26.91 | [134] | |
| $Ti_{0.525}Zr_{0.525}Mn_{0.9}Fe_{1.1}$ | -24.91 | [134] | |
| $Ti_{0.525}Zr_{0.525}Mn_1Fe_1$ | -25.13 | [134] | |
| $Ti_{0.525}Zr_{0.525}Mn_1V_{0.05}Fe_{0.95}$ | -26.42 | [134] | |
| $Ti_{0.24}Zr_{0.17}Mn_{0.17}Co_{0.17}V_{0.17}Fe_{0.08}$ | -34.1 | [135] | |